\documentstyle[preprint,aps]{revtex}

\tightenlines
\newcommand{\fr}[2]{\frac{\displaystyle{#1}}{\displaystyle{#2}}}
\newcommand{\cfr}[1]{(\ref{#1})}
\newcommand{\cA}{{\cal A}}
\newcommand{\cL}{{\cal L}}
\newcommand{\cP}{{\cal P}}

\input{psfig} 

\begin{document}
\draft
\title{\bf Entropy Production : From Open Volume Preserving to
Dissipative Systems}
\author{T. Gilbert and J. R. Dorfman}
\address{Department of Physics and Institute for Physical Science and
Technology, \\ University of Maryland\\College Park MD, 20742, USA}
\date{\today}
\maketitle
\renewcommand{\thefootnote}{\fnsymbol{footnote}}
\setcounter{footnote}{- 1}
\footnote{E-{\it mail} addresses : {\it tmg@ipst.umd.edu, jrd@ipst.umd.edu}}
\renewcommand{\thefootnote}{\arabic{footnote}}

\begin{abstract}
We generalize Gaspard's method for computing the $\varepsilon$-entropy production rate in Hamiltonian systems to 
dissipative systems with attractors considered earlier by T\'el, Vollmer, and Breymann. This
approach leads to a natural definition of a coarse grained Gibbs entropy
which is extensive, and which can be expressed in terms of the SRB
measures and volumes of the coarse graining sets which cover the attractor. One can also study
the entropy and entropy production as functions of the degree of resolution of the coarse
graining process, and examine the limit as the coarse graining size
approaches zero. We show that this definition of the
Gibbs entropy leads to a positive rate of irreversible entropy
production for reversible dissipative systems. We apply the 
method to the case of a two
dimensional map, based upon a model considered by Vollmer, T\'el
and Breymann, that is a deterministic version of a
biased-random walk. We treat both volume preserving and
dissipative versions of the basic map, and make a comparison between
the two cases.  We discuss the $\varepsilon$-entropy production rate
as a function of the size of the coarse graining cells for these
biased-random walks and, for an open system with flux boundary conditions, 
show regions of exponential growth and decay of the 
rate of entropy production as the size of
the cells decreases. 
This work describes in some detail the relation between the results of 
Gaspard, those of of T\'el, Vollmer and Breymann, and those of Ruelle, on 
entropy production in various systems described by Anosov or Anosov-like 
maps. 
\end{abstract}
\section{Introduction}

Over the past few years, a great deal of attention has been devoted to
the issues of entropy production in chaotic, thermostated
systems subjected to external fields, 
\cite{hoo1,evmo,cels,rue1,btv}. 
Such systems are often used in molecular dynamics simulations
of irreversible processes in fluids, such as shear flows, or diffusive
flows. The external field is used to provide a mechanism to establish
a flow in the system, and the thermostat maintains a
constant kinetic or total energy in the system, and produces a
non-equilibrium, stationary state. The presence of the thermostat is felt in the dynamics of the
particles of the system, which becomes, in the usual configuration and momentum
variables, at least, a non-Hamiltonian, non-symplectic system \cite{detmor}. 
The theoretical analyses of these thermostated systems has led to very
interesting and fruitful connections between chaotic dynamics,
transport coefficients, and irreversible thermodynamics, 
\cite{holhoo,hoover,poshoo,ecm1,ecm2,bec,gallco,romo,dorvbei,dorfman}.   

One of the results of this analysis is a relation between transport
coefficients, such as the coefficient of shear viscosity or of
diffusion, and the sum of all of the Lyapunov exponents of the
system, \cite{cels,ecm1}. This sum, in contrast with that of a Hamiltonian, 
symplectic system, is not zero but is negative, and is proportional to the 
square of the external
field strength, for small enough external fields. This connection is
generally established by means of entropy production arguments,
whereby two expressions for the irreversible entropy production in a
thermostated system are set equal to each other. One of these
expressions is just the usual relation between the irreversible
entropy production per unit time, $\sigma$, and the fluxes, $J_{i}$ and forces, $X_{i}$, in an irreversible
process, given by
\begin{equation}
\sigma =\sum_{i}J_{i}X_{i}=\sum_{i,j}L_{ij}X_{i}X_{j}.
\label{neq1}
\end{equation}
Here we assumed that, for thermostated systems in small external fields, the fluxes, $J_{i}$,
are related to the forces, $X_{j}$, through linear laws
\begin{equation}
J_{i}=\sum_{j}L_{ij}X_{j}.
\label{neq1a}
\end{equation}
In Eqs. (\ref{neq1},\ref{neq1a}), the quantities $L_{ij}$ are the Onsager
coefficients, which are directly related to the transport
coefficients, and are supposed to form a positive definite matrix, so
that the entropy production per unit time is positive \cite{degmaz}. 
The other of these two expressions for the entropy production is rather
problematic. Usually one considers the time derivative of the Gibbs
entropy, $S_{G}$, for the thermostated system, given by
\begin{equation}
\frac{d}{dt}S_{G}(t) \equiv -\frac{d}{dt}\int \, d\Gamma
\rho(\Gamma,t)[\ln \rho(\Gamma,t)-1].
\label{neq2}
\end{equation}
Here $\Gamma=(q_{1}, ...,q_{Nd},p_{1},...,p_{Nd})$ is a point in the phase space of the system, and
$\rho(\Gamma,t)$ is the phase space density of the system. Here the
$q_{i},p_{i}$, for $ 1\leq i \leq Nd,$ are the configuration and momentum
variables of a system of $N$ particles in $d$ space dimensions. Because
the system is no longer Hamiltonian, $\rho$ no longer obeys the
Liouville equation. However, $\rho$ does obey a conservation equation of
the form
\begin{equation} 
\frac{\partial}{\partial t}\rho(\Gamma,t) =  
-\sum_{i}\left[\frac{\partial}{\partial
q_{i}}(\dot{q}_{i}\rho(\Gamma,t))
 +\frac{\partial}{\partial p_{i}}(\dot{p_{i}}\rho(\Gamma,t))\right],
\label{neq3}
\end{equation}
with the usual ``dot" notation for the time derivative of a dynamical
variable. This conservation equation can also be expressed in terms of
the total time derivative of $\rho(\Gamma,t)$ as
\begin{equation}
\frac{d\rho}{dt}
=-\rho(\Gamma,t)\sum_{i}\left[\frac{\partial\dot{q_{i}}}{\partial
q_{i}}+\frac{\partial\dot{p_{i}}}{\partial p_{i}}\right].
\label{neq3a}
\end{equation}
We will use this form below.

As is well known, for a closed, Hamiltonian system, or one with
periodic boundary conditions, the Gibbs entropy is
constant in time, but for a thermostated system this is no longer
true. Instead one finds, after using Eq. (\ref{neq3a}), and some partial
integrations \cite{cels}, that
\begin{equation}
\frac{d}{dt}S_{G}(t) =\int d\Gamma
\rho(\Gamma,t)\sum_{i}\left[\frac{\partial\dot{q}_{i}}{\partial
q_{i}}+\frac{\partial\dot{p}_{i}}{\partial p_{i}}\right].
\label{neq4}
\end{equation}
Here we have assumed
that the phase space distribution function vanishes at all boundaries
in configuration and momentum space. For a Hamiltonian system, the right hand
side of Eq. (\ref{neq4}) vanishes, but for a
thermostated system, the divergence of the phase space velocity is
not zero.  The right hand side of Eq. (\ref{neq4}) is easily
related to the sum of the Lyapunov exponents of the system, by the
following argument. We write the phase space density $\rho(\Gamma,t)$
as 
\begin{equation}
\rho(\Gamma,t) =\frac{{\cal N}}{{\cal V}(t)},
\end{equation}
where ${\cal N}$ is a fixed number of members of the ensemble in a
small phase space volume, ${\cal V}(t)$. It then follows from the
conservation equation, Eq. (\ref{neq3a}), that
\begin{equation}
\frac{d\rho}{dt} = -\rho\sum_{i}\left[\frac{\partial \dot{q}_{i}}{\partial
q_{i}}+\frac{\partial \dot{p}_{i}}{\partial p_{i}}\right]=  -\rho\frac{d\ln{\cal V}(t)}{dt}. 
\label{neq4a}
\end{equation}
Now the small phase space volume, ${\cal V}(t)$, changes with time as
\begin{equation}
{\cal V}(t) \simeq {\cal V}(0)\exp[t\sum_{j}\lambda_{j}],
\label{neq4b}
\end{equation}
where $\lambda_{j}$ are the local Lyapunov exponents at the point in
phase space where $\rho$ is evaluated. It then follows immediately
from Eqs. (\ref{neq3a}-\ref{neq4b}), that the rate of entropy
production is given by
\begin{equation}
\frac{d S_{G}(t)}{dt} = \int d\Gamma
\rho(\Gamma,t)\sum_{j}\lambda_{j}(\Gamma) = \sum_{j} <\lambda_j>,
\label{neqc}
\end{equation}
and is negative whenever there is an average contraction of phase space
volumes~! A very similar argument, using the Frobenius-Perron equation, shows that this conclusion is also
valid for maps as well as for flows. This circumstance makes it difficult to equate the positive entropy
production from irreversible thermodynamics to the negative change in
the Gibbs
entropy. 

This paradoxical situation is usually resolved in
the literature, \cite{cels,rue1,dorfman}, by 
saying that the negative entropy production inside
the system is compensated by a positive entropy production in the
thermostat itself, so that the total entropy production per unit time of the 
[system + thermostat] is positive, or  zero. One assumes that a
non-equilibrium steady state is produced, eventually, in which the
total entropy production in the [system + thermostat] is zero. Then the
hypothetical (positive) entropy
production in the thermostat is equated to the
phenomenological (positive) entropy production, which produces the desired
relation between the Lyapunov exponents and the transport
coefficients. This procedure is not quite satisfactory for the following
reasons: 

\par (1) One expects from
phenomenological arguments that
the rate of change of the local entropy, $s(\vec{r})$, in a small region about a point $\vec{r}$ can be decomposed into a term that represents the entropy flow into or out of the region plus a term that represents the local irreversible entropy production within the region. For thermostated systems one would like to represent the entropy flow term as the sum of two pieces, one representing the flow of entropy from neighboring regions due to physical currents, and another term representing the flow of entropy to or from the thermostat \cite{vtb}.
This suggests that, for thermostated
systems, the local entropy change should be written as
\begin{equation}
\frac{d s}{dt} = \frac{d_{th}s}{dt} + \frac{d_{e}s}{dt} + \frac{d_{i}s}{dt},
\label{phen}
\end{equation}
where the first term on the right hand side is identified with the local flow
of entropy from the thermostat to the region.
The second term, denoted by the subscript $e$, is the rate of flow of
entropy into the region from its local environment, and the third term, 
denoted by the subscript $i$, is  the local rate of
irreversible entropy production in the region. In the analysis described in 
Eq. (\ref{neq4}) for thermostated systems, 
the thermostat appears in the equations of motion for the particles as a sort 
of dynamical friction which depends upon the phase point of the
particles, and not as a source of a physical current of particles, momentum 
or energy into or from the system at the boundaries. By using continuous 
distribution functions as described above, one finds a negative rate of 
change of the total Gibbs entropy for the system, and one
assumes there is a compensating positive entropy flow to the
thermostat. The positive rate of change in the thermostat is then identified 
with the irreversible entropy production required by the Second Law. In this 
approach however there is no clear indentification of a positive irreversible 
entropy production within the system. Instead, the total entropy change in 
the system is due to the interaction with the thermostat, which lowers the 
entropy of the system, in effect, by reducing the phase space available to 
the system to a fractal attractor of lower information dimension than the 
phase space itself, as discussed below. 

\par (2) This last remark points to an additional, and perhaps deeper, problem
with the analysis
given in Eq. (\ref{neq4}), especially as it is applied to a
system in a non-equilibrium stationary state. If we think of the
approach of a thermostated system to a non-equilibrium steady state,
then it is not surprising that the Gibbs entropy for the system
decreases with time. That is, let us think of a positive entropy
production as a loss of information about the system, and a negative
entropy production as a gain of information about the system. Since the
thermostat has the effect of creating an attractor in phase space, as time 
goes on the phase point of the system gets closer and closer
to the attractor, \cite{holhoo,hoover,poshoo}. Thus we learn more 
and more about the location of
the phase point of the system as time increases, and this gain of
information is reflected in a negative entropy change. However, as
emphasized by by Breymann, T\'el, and Vollmer \cite{btv} and by Gaspard\cite{gas1}, this use of the
Gibbs entropy supposes that we have some way to locate a
phase point with arbitrary precision. Further, we have also assumed,
in computing the rate of change of the Gibbs entropy, that the
distribution function $\rho(\Gamma,t)$ is a differentiable
function. While this may be true as the system evolves toward a steady
state, it is no longer true in the steady state. Instead, for 
Anosov systems\footnote{As pointed out by Gallavotti and Cohen \cite{gallco}, 
the assumption that the system be Anosov can be replaced by Anosov-like,
which assumes that the flow be hyperbolic on a subset of the whole
phase space that differs from it by a set of measure zero. For the purpose
of this paper, the only property that we will require of Anosov-like 
systems is the existence of a generating partition, namely that the dynamics 
can be encoded in an unambiguous way by a sequence of symbols taken from a
finite alphabet.}  with phase 
space contraction, the phase
space distribution becomes a singular SRB measure on the attractor with a
smooth distribution in the unstable directions and a fractal structure
in the stable directions. This type of distribution precludes the use
of the ordinary calculus of differentiable functions, and requires a
more careful analysis of the entropy production in the steady
state. Therefore one cannot use differentiable functions to describe
the distribution function for the system in a non-equilibrium steady
state, and the calculation leading to Eq. (\ref{neq4}) is not correct.

Gaspard \cite{gas1}, in a study of the entropy production in open, 
Hamiltonian
systems, provided a method for analyzing the entropy production in a
system with a phase space distribution function that is, properly
speaking, a singular measure. He considered a two-dimensional multi-baker map,
constructed so as to allow diffusion of non-interacting particles in one 
space dimension. He
then placed a high density reservoir at one end of the multi-baker
chain, and a low density reservoir at the other end of the chain. The
map then sets up a steady state in which a non-uniform density profile
is established along the chain, and the phase space distribution for a
particle in the chain, in the infinite system limit, is a nowhere 
differentiable SRB measure, smooth in the unstable
direction, but fractal in the stable direction, see also
\cite{tasgas}. Since the properties of the SRB measure rule out the
use of differentiable distribution functions,  
Gaspard 
constructed a so-called $\varepsilon$-entropy, which can be thought of
as a kind of coarse grained entropy, appropriate for singular,
non-differentiable measures, provided the rate of entropy production is 
reasonably insensitive to the size of the coarse graining regions. Further, 
this system has no thermostat, and there is a clear separation of the local
$\varepsilon$-entropy change into a local $\varepsilon$-entropy flow and a 
local irreversible $\varepsilon$-entropy production. The latter is positive, 
and for large systems, it depends on the density gradient in a way that agrees
precisely with the laws of irreversible thermodynamics. 

Similarly, Vollmer {\it et al.} \cite{vtb} considered a version of the 
multi-baker map which, though time reversible, is not volume preserving. They
argued that their model has the same feature as one sees in
thermostated systems, namely the contraction of the
distribution function onto an attractor, and they used it to model the entropy
production in a thermostated system. They  
considered the coarse grained entropy production in diffusive
processes taking place in this system. Their analysis was based on a
coarse graining procedure which uses larger coarse graining regions
than that used by Gaspard \cite{gas1}, but the results were very much the 
same. A positive production of the entropy was found and the form of this
entropy production, in an appropriate macroscopic limit, agrees with the result
one expects from nonequilibrium thermodynamics. Moreover, by comparing
the results obtained for a volume preserving version of their model
with those of the volume non-preserving version, they identified the
effect of the thermostat on the rate of entropy production by looking at
the difference between the rates of entropy production in the two versions.

\par What is perhaps most novel in the 
work of Gaspard and of Vollmer {\it et al.} is the
fact that one now has a firm physical and mathematical reason for the
coarse graining, namely the existence of underlying singular measures
in phase space. Fractal structures in phase space also appear as the
support for physical measures for Hamiltonian systems as
well. The proper treatment of these phase space measures requires the
use of coarse graining methods in more general cases than those
considered here, of course. Moreover, in the proper description of many-particle
systems, the use of coarse graining methods arise naturally when
one goes from the Gibbs $\Gamma$-space description of the system, with
a zero fine grained rate of entropy production, to the Boltzmann
$\mu$-space description involving reduced distribution
functions\cite{tolman,ehrenfest}. However it requires an understanding of 
the hyperbolic nature of the many-particle system, and the structures in 
phase space along stable and unstable directions to understand why the
reduced distribution functions, themselves, approach their equilibrium
values in time, with a positive generation of entropy in the process
\cite{dorfman}.  
 
\par To summarize~: For both volume preserving systems driven out of
equilibrium and volume non-preserving systems whose microscopic dynamics are 
described by Anosov-like systems, one should use a coarse grained entropy and
entropy production to properly describe the system in a
nonequilibrium steady state, since the
distribution function is not smooth on any scale, no matter how
fine. The use of a coarse grained entropy automatically involves a
loss of information about the system since there is always a level of
detail about the system which is inaccessible to the coarse grained
description. This loss of information can then be identified with a
positive irreversible entropy production. In the examples studied
so far, this positive entropy production agrees, in the proper limit,
with the predictions of irreversible thermodynamics. This agreement is
ultimately the chief requirement of any microscopic definition of
entropy production. Once one has a good microscopic definition of the rate of
entropy production in a thermostated system, one can then try to
identify the various terms in Eq. (\ref{phen}) with their microscopic
counterparts. We are then provided with a means to identify the role
of the thermostat in the production of entropy of the system. 

\par The purpose of this paper is to present a unified view of positive entropy
production in both reversible, volume preserving and in reversible,
volume ``contracting" maps, which are time-discretized versions of
thermostated dynamical systems. This paper can be
considered to be a commentary upon and elaboration of work by Gaspard 
\cite{gas1} and by Vollmer, T\'el and Breymann \cite{vtb}. 
What is new here is the observation
that Gaspard's $\varepsilon$-entropy procedure, when generalized appropriately, can be usefully applied
to both Hamiltonian and thermostated systems, and that for the latter
systems, there is much to be gained by a study of the effects of
coarse graining on arbitrarily fine scales. In particular, we are able
to relate the rate of irreversible entropy production to the difference of the
entropies at two levels of resolution of the phase space, (see Eq. (\ref{lk_ent_prod})), and for simple
models we can explicitly evaluate the entropy production as a function
of the level of resolution of the coarse graining procedure. Moreover,
by carefully studying the effects of different levels of resolution,
we can review and refine the relations described above between entropy
production, Lyapunov exponents, and transport coefficients. This will
allow us to make contact with recent work of Chernov {\it et al.} \cite{cels}
and Ruelle \cite{rue1}, establishing rigorous results on the rate of entropy 
production in thermostated systems. Vollmer, T\'el, and Breymann
\cite{vtbpre} have recently done related and complementary
work on the effects of coarse graining in multi-baker models and on the 
relation of their work with that of Gaspard, Chernov {\it et al.}, and 
Ruelle. However, they do not
look at the effects of making the coarse graining cells arbitrarily
small, as we do here. We should also mention that while our work here does shed some further light on the nature of irreversible entropy production in simple thermostated systems, it does not answer a host of questions about the nature of entropy production in the more general setting of a many particle system with a large number of degrees of freedom. For such systems our results may be relevant for examining the rate of entropy production when the phase space distribution is projected onto a subspace of a few relevant variables.

In the next section, we define a coarse grained, local entropy which is
extensive and depends upon both the measure and the volume of each of the coarse
graining regions in phase space. In section III we describe the rate
of change of this coarse grained entropy in a nonequilibrium
stationary state. We show that the rate of change of this local entropy is
in fact zero in the steady state, but that it can be further
decomposed in a way that is consistent with Eq. (\ref{phen})
for dynamical systems with 
generating partitions, such as a Markov partition. In particular, this
applies to  those systems that satisfy Gallavotti and Cohen's 
chaotic hypothesis \cite{gallco}. For systems where there is no flow of 
particles through the boundaries, and the distribution function vanishes at 
the boundaries, we derive, in section IV, a positive irreversible entropy 
production rate which is equal to the phase space contraction rate, in 
agreement with Chernov {\it et al.} \cite{cels} and Ruelle \cite{rue1}. 

\par The general method discussed
here is then applied to both volume preserving and volume contracting
multi-baker chains, which represent deterministic models of biased
random walks on a line. We show, in particular, that our definition of
the entropy production leads to results consistent with those of Gaspard
\cite{gas1}, and with Vollmer {\it et al.} \cite{vtb}, which 
we generalize to arbitrary resolution parameters. In particular, we 
show that, in its leading order, the entropy production rate for these
maps increases 
exponentially as a function of the resolution parameter for a range of 
resolution parameters, and then, as the finite size effects start to 
interfere, falls exponentially to zero as the resolution gets finer and finer. 
We conclude with a discussion of a number of points
raised by this work.

\section{Gibbs entropy for contracting systems}

\par We begin by considering a dynamical system defined by a map $\Phi$ on a 
phase space $X$ with invariant measure $\mu.$ As discussed above, the Gibbs 
entropy for this system, if it were to be well-defined, would be given by 
Eq. (\ref{neq2}), as
\begin{equation}\label{Gibbs}
S_G = - \int_X d\Gamma \rho(\Gamma)[\ln(\rho(\Gamma))-1],
\end{equation}
where the phase space density, $\rho(\Gamma)$, at a point $\Gamma$,
would be the derivative of the measure of a small region about
$\Gamma$, 
\begin{equation}\label{rn_der}
\rho(\Gamma) = \fr{d\mu(\Gamma)}{d\Gamma}.
\end{equation}
That is, $\rho$ would be the density of $\mu,$ formally the Radon-Nikodym 
derivative of
$\mu$ with respect to the Liouville measure in phase space. However, as
noted above, we must be
cautious here, since the existence of a phase space density is only guaranteed
for measures that are absolutely continuous with respect to the 
Liouville measure. In particular, reversible systems in nonequilibrium
steady states do not usually satisfy this requirement, especially, but not 
exclusively, if the system is dissipative with a contraction of phase space 
volumes.
Therefore we cannot define the Gibbs entropy as in Eq. (\ref{Gibbs}).
Rather, we should define it as
\begin{equation}\label{CG_Gibbs}
S_G = - \int_X \mu(d\Gamma)\left[\ln\fr{\mu(d\Gamma)}{d\Gamma}-1\right].
\end{equation}
\par To give this expression a clear meaning, we assume that our dynamical 
system admits
a {\it generating partition} $\cA$ (see for instance \cite{sinai}) and define 
the $(l, k)$-partition $\cA_{l,k}$
as
$$\cA_{l, k} =  \Phi^{- l}(\cA)\vee\Phi^{- l + 1}(\cA)\vee\cdots\vee
\Phi^{-1}(\cA)\vee\cA\vee\Phi(\cA)\vee\cdots\vee\Phi^{k-2}(\cA)\vee
\Phi^{k-1}(\cA).$$
That is, we suppose that there is some partition, $\cA$ of the phase space 
into small, disjoint sets. We then consider forward iterations of these sets, 
which we denote by $\Phi^{j}(\cA)$, and backward iterations, which we denote 
by $\Phi^{-j}(\cA)$. The collection of very many sets denoted by $\cA_{l,k}$ 
above is obtained by taking all possible intersections of the sets generated 
by iterating $\cA$ forward in time over $k-1$ steps and backwards in time by 
$l$ steps.  
Here we use the standard $\vee$ notation for indicating a partitioning
of sets into the collection of all the possible intersections of all
the indicated sets. For an element $A$ of $\cA_{l, k},$ we further define the 
corresponding
volume
\begin{equation}\label{lk_vol}
\nu(A) = \int_X d\Gamma \chi_A(\Gamma),
\end{equation}
where 
$$\chi_A(\Gamma) = 
\left\{\begin{array}{lr}
1,&\Gamma \in A,\\
0 &{\rm otherwise},
\end{array}\right.$$
is the characteristic function of $A.$
\par We now define the $(l,k)$-entropy of the triplet $(X, \Phi, \mu)$, with
respect to the measures and volumes of the elements of this partition, by
\begin{equation}\label{lk_ent}
S_{l, k}(X) = - \sum_{A\in \cA_{l, k}} \mu(A)\left[\log\fr{\mu(A)}{\nu(A)}
- 1\right].
\end{equation}
where $\mu(A)$ is the steady state SRB measure of the set $A$. With our assumption that the partition $\cA$ be generating, the elements
of $\cA_{l, k}$ shrink to points in the limit when both 
$l, k \rightarrow\infty.$ Hence
\begin{equation}
\lim_{l,k\rightarrow\infty}S_{l, k} = S_G,
\label{sg}
\end{equation}
as defined in Eq. (\ref{CG_Gibbs}). That is, we construct an {\it extensive}
entropy for a particular refinement of our generating partition, and
then define the Gibbs entropy to be the limit of the entropy, as the
sets of the partition become more and more refined. 
From a physical point of view, it is expected that the limit in 
Eq. (\ref{sg}) be independent of the choice of the partition, and this will 
indeed be the case for the case of a generating partition.
In many cases,
including those discussed here, the
limit in Eq. (\ref{sg}) is negative infinity, since the measure of a
set typically decreases more slowly than its volume as the coarse graining
cells become small. This occurs whenever the SRB measure is not
absolutely continuous with respect to the Lebesgue measure, and the
information dimension of the sets are smaller than their phase space
dimension\footnote{We are indebted to P. Gaspard for this
observation.} However, we will see subsequently that the rate of
entropy production remains well defined as the coarse graining size
approaches zero.

There is a subtle limiting procedure being carried out here, that we
wish to explain in more detail. If we allow the system to reach a
non-equilibrium steady state, we can imagine that the measure $\mu$
is an SRB measure which does not have a well defined
density. However the measure and the volume of the elements of the
partition are well defined, as is the entropy function $S_{l,k}$, for
all $l,k>0$. However, in the conventional approach to the
Gibbs entropy for phase space distributions, one always assumes that
the phase space density is well defined, in effect, assuming that the
limit of an infinitesimally fine partition can be taken {\it before}
the limit $t\rightarrow\infty$, and that any non-equilibrium 
quantity based on that will be well defined in the non-equilibrium 
stationary state. The exchange of limiting processes is an
essential feature of the correct treatment of entropy production in
non-equilibrium steady states.

\par Consider, now, a region $B$ of $X$ whose borders coincide with the
borders of some elements of $\cA_{l'k'},$ for some $l'$ and $k'.$ For
all $l > l'$ and $k > k',$ we {\em define} the 
$(l,k)$-entropy of $B \subset X$ with respect to $\mu$ by
\begin{equation}\label{lk_ent_B}
S_{l, k}(B) = - \sum_{A\in \cA_{l, k}\cap B}
\mu(A)\left[\log\fr{\mu(A)}{\nu(A)} - 1\right].
\end{equation}
\par In the next section, we derive the $(l, k)$-entropy change in a time 
dependent picture and compare it to Eq. (\ref{phen}) in order to 
identify the various terms in the rate of change of the local entropy. 

\section{coarse grained entropy change}

\par In a time dependent picture, the evolution of the density $\rho_t$
is given by the action of the Frobenius-Perron operator, $\cP$. For an 
invertible map $\Phi$, we have
\begin{equation}\label{PF}
\rho_{t + 1}(\Gamma) = \cP\rho_t(\Gamma) = 
\left|\fr{d}{d\Gamma}\Phi^{- 1}(\Gamma)\right|\rho_t(\Phi^{-1}(\Gamma)),
\end{equation}
where the derivative in the third term indicates the Jacobian of
$\Phi^{-1}(\Gamma)$ with respect to $\Gamma$. Since the quantity $\rho_{t}$ is well defined for finite $t$, we will use it to construct the measures needed for the computation of the entropy, $S_{l,k}(B)$, and entropy changes. The idea is to express the quantity $\mu(A)/\nu(A)$ appearing in the logarithm on the right hand side of Eq. (\ref{lk_ent_B}) as a coarse grained density by means of the relation
\begin{equation}
\frac{\mu_{t}(A)}{\nu(A)} =\overline\rho_{t}(A) \equiv \fr{1}{\nu(A)}\int_{A}\rho_{t}(\Gamma)d\Gamma.
\label{cgdef}
\end{equation}

\par Let us now apply the Frobenius-Perron equation to the evolution of a coarse grained density 
$\overline\rho_t(A)$ of some set $A$. Using Eq. (\ref{PF}), we have
\begin{eqnarray}
\overline\rho_{t + 1}(A) &\equiv& \fr{1}{\nu(A)}\int_A\rho_{t + 1}(\Gamma)
d\Gamma,\nonumber\\
&=&\fr{1}{\nu(A)}\int_A\left|\fr{d}{d\Gamma}\Phi^{- 1}(\Gamma)\right|
\rho_t(\Phi^{-1}(\Gamma))d\Gamma,\nonumber\\
&=&\fr{1}{\nu(A)}\int_{\Phi^{- 1}(A)}\rho_t(\Gamma)d\Gamma,\nonumber\\
&=&\fr{1}{\nu(A)}\mu_{t}(\Phi^{- 1}(A)).\label{cgdens}
\end{eqnarray}
Note that the definition of the measure $\mu_{t}(A)$, together with the Frobenius-Perron equation implies that
\begin {equation}
\mu_{t+1}(A) = \mu_{t}(\Phi^{-1}(A)).
\label{meas}
\end{equation}
\par Now consider a region $B$ such as discussed earlier. The change of
the $(l, k)$-entropy of
$B$ at time $t$ is given by
\begin{equation}\label{FP_ent_change}
\widetilde{\Delta S}_{l,k} (B, t) = S_{l,k}(B, t + 1) - S_{l,k}(B, t).
\end{equation}
That is, $\widetilde{\Delta S}_{l,k} (B, t)$ is the entropy change of $B$ with 
respect to the $(l,k)$-partition.
Then making use of Eqs. (\ref{cgdef},\ref{cgdens}), we find
\begin{eqnarray}
\widetilde{\Delta S}_{l,k} (B, t) &=& 
\sum_{A\in\cA_{l,k}\cap B}\left[ - \mu_{t + 1}(A)
\log\overline\rho_{t + 1}(A) + \mu_{t}(A)\log\overline\rho_t(A)\right],
\nonumber\\
&=& \sum_{A\in\cA_{l,k}\cap B}\left[ - \mu_{t}(\Phi^{- 1}(A))\log
\fr{\mu_{t}(\Phi^{- 1}(A))}{\nu(A)} + \mu_{t}(A)\log\fr{\mu_{t}(A)}{\nu(A)}\right].
\end{eqnarray}
We now have an expression for the time dependent change in the coarse grained entropy of some set $B$ in phase space. Suppose we keep the $(l,k)$-partition fixed but consider the limit $t\rightarrow\infty$. If the system reaches a nonequilibrium steady state, then we can expect that the limit 
\begin{equation}
\lim_{t\rightarrow\infty}\mu_{t}(A)=\mu(A),
\end{equation}
will exist for all sets, $A$, of the partition, and that the measure will be invariant in the stationary state where $\mu(A)=\mu(\Phi^{-1}(A))$, as implied by Eq. (\ref{meas}).
But in this case, the entropy change defined above is zero for an invariant measure, i.e.
\begin{equation}\label{nil_ent_change}
\widetilde{\Delta S}_{l, k} (B) = 0.
\end{equation}
 Thus we have
defined a coarse grained entropy which has a zero rate of change in
the non-equilibrium steady state. We now have to decompose it into the
three contributions required by Eq. (\ref{phen}). We first consider the time dependent case and then specialize to the case of a steady state with an invariant measure.

\par First, we define the rate of $(l,k)$-entropy {\it flow} into the set $B$ at time $t$,
$\Delta_e S_{l,k}(B,t)$, by taking the
difference between the
$(l,k)$-entropy of the pre-image of $B$, which we take to be the entropy of set $B$ after the next time step, and the $(l,k)$-entropy of $B$,
itself. That is, there is no contribution to the {\em flow} of entropy into $B$ from the set of points which are in $B$  at both $t$ and $t+1$. Thus, the rate of $(l,k)$-entropy flow of $B$ is defined to
be
\begin{equation}\label{lk_ent_flow}
\Delta_e S_{l,k} (B,t) = S_{l, k}(\Phi^{-1}(B),t) - S_{l,k}(B,t).
\end{equation}
 
\par Next we can define the flow of entropy into the set $B$ due to
the presence of the thermostat, $\Delta_{th}S_{l,k}(B,t)$. Since the 
thermostat is modelled by adding frictional terms to the equations of motion 
of the points and not by any boundary conditions,  we have no means of 
identifying the action of the thermostat other than by the change in the 
volume of sets in the course of their time evolution. This change in volume 
is produced by the frictional terms added to the equations of motion. A 
reasonable definition of $\Delta_{th}S_{l,k}(B)$ should satisfy the 
requirement that this
entropy flow vanishes if the transformation $\Phi$ is volume
preserving. This condition is satisfied by {\em defining} the flow of
entropy into $B$ due to the presence of the thermostat by
\begin{eqnarray}
\Delta_{th}S_{l,k}(B,t) & = &
S_{l,k}(B,t+1)-S_{l+1,k-1}(\Phi^{-1}(B),t),\nonumber \\
& = &  - \sum_{A\in\cA_{l,k}\cap B} 
\mu_{t}(\Phi^{-1}(A))\log\fr{\nu(\Phi^{-1}(A))}{\nu(A)}.
\label{sth}
\end{eqnarray}
To obtain the second line in Eq. (\ref{sth}), we have used Eq. (\ref{meas}), and the fact
that the 
pre-images of sets in 
$\cA_{l, k}$ are sets in $\cA_{l + 1, k - 1}$.  We call attention
to the fact that the pre-images of the sets in a $(l,k)$-partition are
sets in a 
$(l+1, k-1)$-partition. We have defined $\Delta_{th}S_{l,k}(B)$ as the
difference between the entropy of the set $B$ at time $t+1$, and the
entropy of the pre-image of $B$ at time $t$, where the entropy of the 
pre-image sets
of $B$ are calculated using the pre-image of the $(l,k)$ partition. This 
definition of the flow of entropy from
the thermostat to the set $B$ satisfies the requirement that it
vanishes for a volume preserving transformation. The definition of 
$\Delta_{th}S_{l,k}(B,t)$ given above suffers from the lack of a clear 
derivation based upon a physical picture of a thermostat, as one would 
expect for a thermostat which acts only at the boundary of the system. Here 
the thermostat is an ``internal'' device which has the effect of modifying 
the equations of motion and produces a contraction of phase space volumes. 
Thus our definition of entropy flow to the system from the thermostat must 
be based on the change of the volumes of cells in phase space with time, as 
is done in our definition above.

We now want to follow the phenomenological approach as in
Eq. (\ref{phen}) and write
\begin{equation}
\widetilde{\Delta S}_{l,k}(B,t)=\Delta_e
S_{l,k}(B,t)+\Delta_{th}S_{l,k}(B,t)+\Delta_{i}S_{l,k}(B,t),
\label{irenp}
\end{equation}
where $\Delta_{i}S_{l,k}(B,t)$ represents the rate of irreversible entropy
production in $B$. 
By combining Eqs. (\ref{irenp}), with Eqs. (\ref{lk_ent_flow}) and
(\ref{sth}), we obtain an expression for the irreversible entropy
production in set $B$ as   
\begin{equation}\label{lk_ent_prod}
\Delta_i S_{l,k} (B,t) = S_{l + 1, k - 1}(\Phi^{-1}(B),t) -
S_{l,k}(\Phi^{-1}(B),t).
\end{equation}
Note that this equation represents the irreversible entropy production in 
any set $B$ as the difference in the entropy of the pre-image sets at two 
levels of
resolution. This is an important result which helps us to understand
that irreversible entropy production is a direct result of the loss of information
in the coarse graining of a system.

\par Therefore, we have been able to use the phenomenological approach to
entropy production, Eq. (\ref{phen}) and some reasonable definitions
of entropy flows to obtain an expression for the local rate of
irreversible entropy production.

\par We can easily connect
these definitions to that of Gaspard \cite{gas1} for the entropy production in
a Hamiltonian system, by defining an {\em intrinsic} local rate of entropy 
change in the set $B$, $\Delta S_{l,k}(B,t)$, by removing the term due to the
thermostat as
\begin{eqnarray}\label{ent_change_redef}
\Delta S_{l,k}(B,t)& = & \widetilde{\Delta S}_{l,k}(B,t) - \Delta_{th}S_{l,k} (B,t), \nonumber \\
& = & S_{l+1,k-1}(\Phi^{-1}(B),t)-S_{l,k}(B,t).
\end{eqnarray}  
That is, we have expressed the intrinsic rate of change of the coarse
grained entropy of $B$  as the difference between the $(l+1,k-1)$-entropy
of the pre-image of $B$ and the $(l,k)$-entropy of $B$ itself. This
expression is identical with the definition of the change in the
coarse grained entropy given by Gaspard \cite{gas1}. Note that in a
non-equilibrium 
steady state, where $\widetilde{\Delta S}_{l,k}(B)=0$, the rate of change of
this intrinsic entropy is equal to the rate of flow of entropy from
the system to the thermostat, $-\Delta_{th}S_{l,k}(B)$, which as we
will see below is positive if there is an average contaction of the
phase space accessible to the system on to an attractor.

\section{({\it \lowercase{l},\lowercase{k}})-Entropy change and phase space contraction}

\par It is important to note that in the steady state, the measure $\mu(A)$ of a set
$A$ is invariant, {\it i.e.}, $\mu(\Phi^{-1}(A))=\mu(A)$. Now, even
though the measure $\mu(A)$ of a set $A$ is invariant, the various
terms in the rate of entropy production may not be zero because the phase space volumes of
the elements of the partition are not equal to the volumes of
their pre-image sets. Thus, using the
fact that the pre-images of sets in $\cA_{l, k}$ are sets in 
$\cA_{l + 1, k - 1}$, we find that the intrinsic rate of
entropy change, $\Delta S_{l,k}(B)$, Eq. (\ref{ent_change_redef}), becomes
\begin{eqnarray}
\Delta S_{l,k}(B) &=& S_{l + 1, k - 1}(\Phi^{-1}(B)) - S_{l,k}(B),\nonumber \\
&=& - \sum_{A\in \cA_{l, k}\cap B}\left[ \mu(\Phi^{- 1}(A))\ln\fr{\mu(\Phi^{- 1}(A))}
{\nu(\Phi^{- 1}(A))} - \mu(A)\ln\fr{\mu(A)}{\nu(A)}\right].
\label{ec1}
\end{eqnarray}
Using the invariance of the measure, we have
\begin{equation}
\Delta S_{l,k}(B) = \sum_{A\in \cA_{l, k}\cap B} \mu(A) \ln\fr{\nu(\Phi^{- 1}(A))}
{\nu(A)},
\label{ec2}
\end{equation}
as noted at the end of the previous section.

\par We also have an expression for the volume of the pre-image sets in
terms of the Jacobian, $J(\Phi^{-1}(\Gamma))
\equiv |d\Gamma/d\Phi^{-1}(\Gamma)|$, of the transformation
\begin{eqnarray}
\nu(\Phi^{- 1}(A))&=&\int_{\Phi^{- 1}(A)}d\Gamma, \nonumber \\
&=&\int_A\fr{d\Gamma}{J(\Phi^{-1}(\Gamma))}, \nonumber \\
&=&\fr{1}{J(\Phi^{-1}(\Gamma_A))}\nu(A),
\label{ec3}
\end{eqnarray}
where we the last line then follows from the mean value theorem and $\Gamma_A$ denotes
an appropriate point in $A.$

Thus
\begin{equation}
\Delta S_{l,k}(B) = \sum_{A\in \cA_{l, k}\cap B} \mu(A) 
\ln\fr{1}{J(\Phi^{-1}(\Gamma_A))}.
\label{ec4}
\end{equation}
In the limit where $l,k\rightarrow\infty$ the sum becomes an integral
over phase space and we find 
\begin{equation}\label{ent_change1}
\lim_{l,k\rightarrow\infty} \Delta S_{l,k}(B) =  \int_{\Phi^{-1}(B)}
\mu(d\Gamma)\ln\fr{1}{J(\Gamma)}.
\end{equation}
Up to a sign change, this result, or, more precisely,  its generalization 
given below,  forms the starting point for Ruelle's analysis of the
entropy production for diffeomorphisms \cite{rue1}. The sign difference is 
due to
Ruelle's starting with the argument that the negative change in the
Gibbs entropy of the thermostated system is compensated by a positive
entropy production. Here we avoid that procedure and
see that it is possible to define a coarse grained entropy which has a
positive rate of change for the system itself. 
\par We are now in a position to prove our main result, i.e. that the intrinsic 
rate of change of the entropy defined by Eq. (\ref{ent_change_redef}) is positive and 
equals the phase space contraction rate, for a closed, thermostated system. 
To do this, we identify the set $B$ with the entire phase space, $X$, and 
notice that, because the system is closed, there is no flow of entropy into 
or out of the system so that $\Delta_e S_{l,k}(X)$
vanishes. We can thus identify the
rate of change of the, now well defined, Gibbs entropy as the
irreversible entropy production in the system, and it is given as
\begin{eqnarray}
\Delta_i S_{G} & = & \lim_{l,k\rightarrow\infty}\Delta S_{l,k}(X) = \int_{X}
\mu(d\Gamma)\ln\fr{1}{J(\Gamma)}, \nonumber \\
&=& - \sum_i \lambda_i,
\label{ec6}
\end{eqnarray}
where the integral in the first line of Eq. (\ref{ec6}) defines the sum
over all of the Lyapunov exponents of the system. Ruelle has proved
that this quantity is positive if the map is a diffeomorphism, and
$\mu$ an SRB measure on $X$, singular with respect to the Liouville measure.
\par This result shows that, for a contracting  system for which the sum
of the Lyapunov exponents is negative, the stationary state intrinsic entropy change,
as defined in Eq. (\ref{ent_change_redef}), 
is positive. It is in exact agreement with what one
expects for the stationary state entropy production rate in these systems.
Moreover, it clarifies the paradox, discussed earlier, that the 
``fine grained''
Gibbs entropy yields negative stationary state entropy change rate, 
\cite{cels,rue1,hoover,dorfman}.

\par In the next sections, we will use the construction of the
$(l,k)$-entropy given above
to compute the $(l,k)$-entropy flow and irreversible entropy production
for two specific cases and show how our formalism corresponds to the results previously obtained by Gaspard \cite{gas1} for 
Hamiltonian-like, volume preserving
maps, and those of Vollmer, T\'el and Breymann \cite{vtb} for
dissipative, volume contracting maps, for open systems with diffusive flows. 

\section{A deterministic biased random walk}

\par We consider a one-dimensional random walk on a lattice where
a particle hops with probability $s$ (resp. $1 - s$) to the right (resp.
left). The diffusion coefficient for this process can be computed from
the Green-Kubo formula, see for instance \cite{dorfman},
\begin{equation}\label{dif_coef}
D = \lim_{T\rightarrow\infty}\fr{\left<(x_T - <x_T>)^2\right>}{2T} 
= 2s(1 - s),
\end{equation}
where $x_T$ denotes the displacement of a random walker after $T$ time
steps.
We can also compute the mean drift velocity of this process,
\begin{equation}\label{vdrift}
v = 1 - 2s,
\end{equation}
measured positively towards the left direction.
\par A reversible deterministic model of this process is the generalized
multi-baker chain defined on the ``phase space''\footnote{In practice, and for 
the sake of definiteness of the measure, we will restrict the map to a 
finite region
$\cL\in{\bf Z}$ and will specify some boundary conditions} ${\bf Z}\times [0, 1]^2$, consisting of a horizontal chain of unit squares. A phase point is labelled by an interger index $n\in{\bf Z}$ and by internal coordinates $(x,y)$ within a unit square. The dynamics of this multi-baker map is given by 
\begin{equation}\label{GMB}
\Phi(n, (x, y)) = \left\{\begin{array}{lr}
\left(n + 1, (\fr{x} {s}, s y)\right),&0\leq x< s,\\
\left(n - 1, (\fr{x - s} {1 - s}, s + (1 - s)y)\right),&s\leq x <1.
\end{array}\right.
\end{equation}
See Fig. (1). This map is volume preserving. 
If we specifically use periodic
boundary conditions and allow no escape of particles from the system,
it is easy to show that the two Lyapunov exponents for this map are
\begin{equation}
\lambda_+ = - \lambda_- = -s\ln s - (1 - s)\ln (1 - s).
\label{le1}
\end{equation}
Notice that for the volume preserving, closed system, the sum of the
Lyapunov exponents is zero.

Alternatively, we can think of the bias in the random walk as being driven by 
the action of some external field and model this driven process by a map that 
is not area
preserving, \cite{vtb}. This would then be a time-discretized model that has
features similar to those of a system in an external field with an
energy preserving thermostat. That is, as pointed out by Vollmer {\it et
al.} \cite{vtb} one hopes to capture in this
model, the effects of a dynamics that contracts, on the average, volumes in
phase space. We thus let
\begin{equation}\label{DMB}
\Phi_c(n, (x, y)) = \left\{\begin{array}{lr}
\left(n + 1, (\fr{x} {s}, (1 - s) y)\right),&0\leq x< s,\\
\left(n - 1, (\fr{x - s} {1 - s}, 1 - s + s y)\right),&s\leq x <1.
\end{array}\right.
\end{equation}
See Fig. (2). The phase space volumes are not locally preserved and
the periodic version of this map has two Lyapunov exponents
\begin{eqnarray}
\lambda_+ & = & -s\ln s - (1 - s)\ln (1 - s),\nonumber \\
\lambda_- & = & s\ln (1 - s) + (1 - s)\ln s.
\label{le2}
\end{eqnarray}
The negative of the sum of the Lyapunov exponents,
\begin{equation}\label{ps_cr}
- (\lambda_+ + \lambda_-) = (2s - 1)\ln\fr{s}{1 - s} > 0,
\end{equation}
is the phase space contraction rate. In the following discussions, we
consider both of these maps and will 
make explicit distinctions when appropriate. When we wish to refer to
the two maps without distinguishing between them, we will use the
notation $\Phi_{(c)}$ for the maps.

\par For either case, we denote by $\mu_n$ the cumulative, or total,  measure 
of the $n$-th unit square. Using the
Perron-Frobenius equation or other methods, \cite{gas1,tasgas}, one can 
easily see that
for either version of the map, $\Phi_{(c)}$, the stationary measure of the 
chain satisfies the equation
\begin{equation}\label{ss}
\mu_n = s\mu_{n - 1} + (1 - s)\mu_{n + 1}.$$
\end{equation}
The solution of this equation is easily found to be
\begin{equation}\label{ss_GMB}
\mu_n = A\alpha^n + B,
\end{equation}
where $\alpha = s/(1 - s)$ and $A$ and $B$ are fixed by the boundary 
conditions. 

\section{({\it \lowercase{l},\lowercase{k}})-entropy and entropy 
production rate for multibakers}

\par We now follow the procedure outlined in Section II, and construct
the $(l,k)$-partitions, and entropies of the maps $\Phi_{(c)}$.

\par The $(l, k)$-partition is a collection of $2^{l + k}$ non 
intersecting rectangles that cover each of the unit squares in the
chain, with an identical covering for each square. Then for each square, the number of
elements along a line in the expanding direction is  $2^l$, and $2^k$
is the number of those 
along the contracting direction. To make this more precise, we introduce 
a symbolic dynamics on the squares. Let us consider one particular
square and notice that any $(l,k)$-partition of that square is generated by 
images or pre-images of the two sets
$$\begin{array}{lcl}
\Gamma(0) &=& \{(x,y)|\: 0\leq y <s\},\\
\Gamma(1) &=& \{(x,y)|\: s\leq y <1\},
\end{array}
$$
for the case of $\Phi$, \cfr{GMB}, and
$$\begin{array}{lcl}
\Gamma_c(0) &=& \{(x,y)|\: 0\leq y < 1 - s\},\\
\Gamma_c(1) &=& \{(x,y)|\: 1 - s\leq y <1\},
\end{array}
$$
for the case of $\Phi_c$, \cfr{DMB}\footnote{Following the notations 
introduced earlier, we have
$$\cA = \{\Gamma_{(c)}(n, \omega_n),\:n\in\cL,\:\omega_n\in\{0,1\}\},$$
where we indexed the elements of the partition by the square $n$ they
belong to. $\cA_{l,k}$ can be generated by taking images and pre-images
of this partition. However, in order for us not to worry at this point
about technicalities involving the boundary conditions, we will find it
more convenient to define the $(l,k)$-partition by iterations of local
maps, i.e. dropping the index $n$ for that purpose.}.
\par An $(l,k)$-set, $\Gamma(\omega_{-l},\ldots,\omega_{k - 1})$ with 
$\omega_j\in\{0,1\}$, $j = -l,\ldots,k -1$, is the set of points $(x, y)$ 
such that (regardless of the lattice coordinate)
$$\Phi^{- j}(x,y)\in \Gamma(\omega_j),\:j = -l, \ldots, k-1.$$ 
Notice also that 
\begin{equation}\label{pre_lkset}
\Phi^{- 1}\Gamma(\omega_{-l},\ldots,\omega_{k - 1})
=\Gamma(\omega'_{-l - 1},\ldots,\omega'_{k - 2}),
\end{equation}
with $\omega'_j = \omega_{j + 1},$.  This expresses the conjugation between
$\Phi$ and the shift operator on symbolic sequences.
We define $\Gamma_c(\omega_{-l},\ldots,\omega_{k - 1})$ in a similar way and
with the same property. As examples, Figs. (3) and (4) show the $(1,1)$ and 
$(0,2)$-sets of $\Phi$ and $\Phi_c$, respectively.
\par We will use the notation $\mu_n(\omega_{-l},\ldots,\omega_{k - 1})$ to
designate the measure of the corresponding $(l,k)$-set of cell $n$ 
irrespective of which map is being considered. We will further denote the 
volume of the corresponding sets by 
$\nu(\Gamma(\omega_{-l},\ldots,\omega_{k - 1}))$ and 
$\nu(\Gamma_c(\omega_{-l},\ldots,\omega_{k - 1}))$ and will use the notation
$\nu(\omega_{-l},\ldots,\omega_{k - 1})$ when we want to avoid referring to
a specific choice of map. We have
\begin{equation}\label{ps_vol}
\begin{array}{lcl}
\nu(\Gamma(\omega_{-l},\ldots,\omega_{k - 1})) &=& 
\prod_{j = -l}^{k - 1}\nu(\omega_j),\\
\nu(\Gamma_c(\omega_{-l},\ldots,\omega_{k - 1})) &=& 
\prod_{j = -l}^{- 1}\nu(\omega_j)\prod_{j = 0}^{k - 1}\nu^*(\omega_j),
\end{array}
\end{equation}
where 
\begin{equation}\label{nu}
\nu(\omega_j)=\left\{\begin{array}{lr}
s,&\omega_j = 0,\\
1 - s,&\omega_j = 1,
\end{array}\right.
\end{equation}
and
\begin{equation}\label{nu*}
\nu^*(\omega_j)=\left\{\begin{array}{lr}
1 - s,&\omega_j = 0,\\
s,&\omega_j = 1.
\end{array}\right.
\end{equation}
\par We now proceed to evaluate the $(l,k)$-entropy of a site $n$ given by
\begin{equation}\label{lk_ent(n)}
S_{l,k}(n) = - \sum_{\omega_{-l},\ldots,\omega_{k - 1}} 
\mu_n(\omega_{-l},\ldots,\omega_{k - 1})\left[\ln
\fr{\mu_n(\omega_{-l},\ldots,\omega_{k - 1})}
{\nu(\omega_{-l},\ldots,\omega_{k - 1})} - 1\right].
\end{equation}
Notice that this is just the coarse grained version of the Gibbs
entropy, defined by Eq. (\ref{lk_ent_B}) where the set $B$ is now the unit
square representing the site $n$. From now on, we will drop the constant 
term in the expression (\ref{lk_ent(n)}) of the $(l,k)$-entropy.
As discussed in the Appendix A, the stationary state 
measure is uniform along the $x$, or expanding, direction. From this
it follows that the entropy is 
extensive with respect to the $x$-direction. Thus,
\begin{equation}\label{ext}
S_{l,k}(n) = S_{0,k}(n),
\end{equation}
and we are allowed to drop the $l$ dependence. When appropriate, we will 
simply refer to the $k$-entropy of $n$ and write $S_k(n).$

\par To derive the $(l,k)$-entropy production rate, use Gaspard's
method \cite{gas1} and write the rate
of intrisic entropy change as the sum of an entropy flow and an irreversible
entropy production, as in Eq. (\ref{ent_change_redef}).
\par To compute the rate of change in entropy, we notice from 
Eq. (\ref{pre_lkset}) 
that the pre-image of an $(l, k)$-set is an $(l + 1, k - 1)$-set. Thus
as in the more general case discussed earlier,
\begin{equation}
\Delta S_{l,k}(n) = S_{l + 1, k - 1}(\Phi_{(c)}^{-1}(n)) - S_{l,k}(n),
\end{equation}
\par Let us take $l = 0.$ The first term on the RHS is then
\begin{eqnarray}
S_{1, k - 1}(\Phi_{(c)}^{-1}(n)) = &- \sum_{\omega_{0},\ldots,\omega_{k - 2}} 
\left[\mu_{n-1}(0, \omega_{0},\ldots,\omega_{k - 2})\ln
\fr{\mu_{n-1}(0, \omega_{0},\ldots,\omega_{k - 2})}
{\nu(0, \omega_{0},\ldots,\omega_{k - 2})}\right. \nonumber \\
&\left.+\mu_{n+1}(1, \omega_{0},\ldots,\omega_{k - 2})\ln
\fr{\mu_{n+1}(1, \omega_{0},\ldots,\omega_{k - 2})}
{\nu(1, \omega_{0},\ldots,\omega_{k - 2})}\right].\label{44}
\end{eqnarray}
But, from Eq. (\ref{ps_vol}), we know that the measures and volumes
appearing in the above equation are given by
\begin{eqnarray}
\mu_{n-1}(0, \omega_{0},\ldots,\omega_{k - 2})
& = & s \mu_{n-1}(\omega_{0},\ldots,\omega_{k - 2}), \nonumber \\
\mu_{n+1}(1, \omega_{0},\ldots,\omega_{k - 2})
& = & (1 - s) \mu_{n+1}(\omega_{0},\ldots,\omega_{k - 2}),\nonumber \\
\nu(0, \omega_{0},\ldots,\omega_{k - 2})
& = & s \nu(\omega_{0},\ldots,\omega_{k - 2}),\nonumber \\
\nu(1, \omega_{0},\ldots,\omega_{k - 2})
& = & (1 - s) \nu(\omega_{0},\ldots,\omega_{k - 2}).
\end{eqnarray}
We can thus rewrite Eq. (\ref{44}) as
\begin{eqnarray}
S_{1, k - 1}(\Phi_{(c)}^{-1}(n)) &=& - \sum_{\omega_{0},\ldots,\omega_{k - 2}} 
\left[s\mu_{n-1}(\omega_{0},\ldots,\omega_{k - 2})\ln
\fr{\mu_{n-1}(\omega_{0},\ldots,\omega_{k - 2})}
{\nu(\omega_{0},\ldots,\omega_{k - 2})}\right. \nonumber \\
&&\left.+(1 - s)\mu_{n+1}(\omega_{0},\ldots,\omega_{k - 2})\ln
\fr{\mu_{n+1}(\omega_{0},\ldots,\omega_{k - 2})}
{\nu(\omega_{0},\ldots,\omega_{k - 2})}\right],\nonumber\\
&=&s S_{0, k - 1}(n - 1) + (1 - s) S_{0,k - 1}(n + 1).
\end{eqnarray}
By making use of Eq. (\ref{ext}), we conclude that the rate of entropy
change at site $n$ satisfies the simple difference equation
\begin{equation}\label{ent_ch}
\Delta S_k(n) = s S_{k - 1}(n - 1) + (1 - s) S_{k - 1}(n + 1) - S_k(n).
\end{equation}
\par We now consider the contributions to this entropy change from the
entropy flow and the irreversible entropy production. The entropy flow rate 
is given by
$$\Delta_e S_{l,k}(n) = S_{l,k}(\Phi_{(c)}^{-1}(n)) - S_{l,k}(n).$$
By the same argument as above,
\begin{equation}
S_{1,k}(\Phi_{(c)}^{-1}(n)) = s S_{0,k}(n - 1) + (1 - s) S_{0,k}(n + 1).
\end{equation}
So that, using Eq. (\ref{ext}) again, we have 
\begin{equation}\label{ent_flux}
\Delta_e S_{k}(n) =  s S_{k}(n - 1) + (1 - s) S_{k}(n + 1) - S_k(n).
\end{equation}
\par With Eqs. (\ref{ent_ch}), and (\ref{ent_flux}), we can derive an expression for 
the irreversible entropy production rate
\begin{eqnarray}
\Delta_i S_k(n) &=&\Delta S_k(n) - \Delta_e S_k(n),\nonumber \\
&=&s [S_{k - 1}(n - 1) - S_k(n - 1)] + 
(1 - s) [S_{k - 1}(n + 1) - S_k(n + 1)], \nonumber \\
&=& S_k(n) - S_{k + 1}(n),
\label{ent_prod}
\end{eqnarray}
where the last line is a consequence of Eq. (\ref{ss}).
Using Eq. (\ref{lk_ent(n)}), we can obtain useful expressions for
the irreversible entropy production  as
\begin{equation}\label{ent_prod_exp}
\Delta_i S_k(n) =  \sum_{\omega_{0},\ldots,\omega_{k}} 
\mu_n(\omega_{0},\ldots,\omega_{k})\ln
\fr{\mu_n(\omega_{0},\ldots,\omega_{k})}
{\nu(\omega_{k})\mu_n(\omega_0,\ldots,\omega_{k - 1})},
\end{equation}
for the case of the volume preserving map, $\Phi$, and
\begin{equation}\label{ent_prod_exp_c}
\Delta_i S_k(n) =  \sum_{\omega_{0},\ldots,\omega_{k}} 
\mu_n(\omega_{0},\ldots,\omega_{k})\ln
\fr{\mu_n(\omega_{0},\ldots,\omega_{k})}
{\nu^*(\omega_{k})\mu_n(\omega_0,\ldots,\omega_{k - 1})},
\end{equation}
for the case of the contracting map, $\Phi_c.$
\par Notice that Eq. (\ref{ent_prod}) contains the important result that
\begin{equation}\label{ent_prod_rec}
\Delta_i S_k(n) = s \Delta_i S_{k - 1}(n - 1) + 
(1 - s) \Delta_i S_{k - 1}(n + 1).
\end{equation}
This relation enables us to compute the $k$-entropy production rate
recursively from a knowledge of the $0$-entropy production rate.

\section{entropy production rate for flux boundary conditions}

\par In this section, we specify our study to the case of flux boundary 
conditions. That is, we consider a chain of $L$ sites and impose the boundary 
conditions :
$$\begin{array}{lcl}
\mu_0 &=& 1,\\ 
\mu_{L + 1} &=& L + 2.
\end{array}$$
This way, there is an average gradient of density of 1 per unit cell across 
the system. With these boundary conditions, the constants $A$ and $B$ in 
Eq. (\ref{ss_GMB}) are found 
to be
\begin{equation}\label{fbc_GMB}
\begin{array}{lcl}
A&=&\fr{L + 1}{\alpha^{L + 1} - 1},\\
B&=&1 - A.
\end{array}
\end{equation}
Fig. (5) shows $\mu_n$ for $L = 100$ and parameter values 
$s = 0.45, 0.5, 0.55.$
Notice that, with the exception of $s = 0.5$ which corresponds to
a linear growth, the exponential growth is so steep that the density is 
almost constant on the larger part of the lattice. In the limit when 
$L\rightarrow\infty$, $\mu_n$ becomes a constant and is either $L + 2$ or
$1$ depending on whether $s < 0.5$ or $s > 0.5$ respectively. As we will
see shortly, it is precisely the exponential profile of the density
that is responsible for the divergence of the $k$-entropy production
rate, see Eqs. (\ref{exp_dep1}, \ref{exp_dep2}).

\subsection{Volume preserving case}
Let us now apply the formulae (\ref{ent_prod_exp}, \ref{ent_prod_rec})
to the system with the specific boundary conditions given by 
Eqs. (\ref{ss_GMB}), and (\ref{fbc_GMB}). We first compute the $0$-entropy 
production rate.
\begin{eqnarray}
\Delta_i S_0(n) &=&  \sum_{\omega_{0}}\mu_n(\omega_{0})
\ln\fr{\mu_n(\omega_{0})}{\nu(\omega_{0})\mu_n},\nonumber\\
&=& s\mu_{n - 1}\ln\fr{\mu_{n - 1}}{\mu_n} + 
(1 - s)\mu_{n + 1}\ln\fr{\mu_{n + 1}}{\mu_n},\nonumber\\
&=&s(A\alpha^{n - 1} + B)\ln\fr{A\alpha^{n - 1} + B}{A\alpha^n + B}
+(1 - s)(A\alpha^{n + 1} + B)\ln\fr{A\alpha^{n + 1} + B}{A\alpha^n + B},\nonumber\\
&=&s(A\alpha^{n - 1} + B)\ln\left[1 + \fr{A\alpha^n}{A\alpha^n + B}
(\alpha^{-1} - 1)\right],\nonumber\\
& & + (1 - s)(A\alpha^{n + 1} + B)
\ln\left[1 + \fr{A\alpha^n}{A\alpha^n + B}(\alpha - 1)\right].
\label{0_ent_prod_vp}
\end{eqnarray}

Fig. (6) shows a numerical computation of this quantity. The dependence
on $n$ is exponential with slope $2\ln\alpha$ with the exception of
$s= 0.5$ for which the entropy production goes like $1/n,$ which is the
case considered by Gaspard \cite{gas1}. 
\par If we now assume $L >> n >> 1$, the second terms in the logarithms are
very small so that we can expand the logarithms around 1 and keep 
the leading terms (up to second order). After carefully examining the relative
sizes of the various terms, we find that the irreversible entropy
production is given by
\begin{equation}\label{0_ent}
\Delta_i S_0(n) = \fr{(1 - 2s)^2}{2s(1 - s)}\fr{A^2\alpha^{2n}}{A\alpha^n + B}.
\end{equation}
Now, if we define the discrete gradient of $\mu_n$ with respect to $n$
as a symmetrized finite difference, i.e.
$$\begin{array}{lcl}
\nabla\mu_n &=& \fr{1}{2}[(\mu_{n + 1} - \mu_n) + (\mu_n - \mu_{n - 1})],\\
&=& A\alpha^n\fr{2s - 1}{2s(1 - s)},
\end{array}$$
then the $0$-entropy rate production \cfr{0_ent} becomes
\begin{equation}\label{ep_vp}
\Delta_i S_0(n) = D\fr{(\nabla\mu_n)^2}{\mu_n},
\end{equation}
where $D$ is the diffusion coefficient (\ref{dif_coef}).
\par
With our recurrence relation \cfr{ent_prod_rec}, we can now carry out the
computation of the $k$-entropy production rate for any $k$. In Figs. (7) and
(8), we show the first ten $k$'s for the same lattice of length $L = 100$ 
and $s = 0.45$ and $s = 0.55$ respectively. Notice that these curves display
some $k$ dependence. In Figs. (9) and (10), we show the $k$ dependence of 
the entropy production for the middle site, $n = 50$, of a 100 site chain,
for both small and large $k$.
The $k$-entropy  appears to increase exponentially for the lower part of the 
$k$ range, Fig. (10), and then starts to decay exponentially, Fig (10). 
\par The exponential growth can be understood directly form Eqs. (\ref{ent_prod_rec},\ref{0_ent}). Indeed, let us assume
\begin{equation}\label{exp_dep1}
\Delta_i S_k(n) = 
\fr{(1 - 2s)^2}{2s(1 - s)}\fr{A^2\alpha^{2n}}{A\alpha^n + B}\beta^k.
\end{equation}
Using Eq. (\ref{ent_prod_rec}), we find
\begin{equation}\label{exp_dep2}
\beta = \fr{(1 - s)^3 + s^3}{s(1 -s)}.
\end{equation}
In appendix A, we rederive this $k$ dependence from the knowledge of
the stationary state and Eq. (\ref{ent_prod_exp_c}), see 
Eq. (\ref{DiSk4}). Notice that this ignores the finite size effects.
Of course, this exponential divergence is of unphysical nature and one
would expect that the entropy production be independent of $k$. In their
model with a third middle band, Vollmer {\it et al.} \cite{vtb} introduced a 
special scaling that allowed them to get rid of this divergence, while keeping
the zeroth order term in Eq. (\ref{exp_dep1}). 
\par The finite size effects are those responsible for the exponential decay at large values of $k$.
To understand these finite size effects, notice that
the value of $k$ gives the number of time steps for which we know where
the points located in a specific set will go. The only possibility for
these sets to produce entropy is if they remain in the chain for more
than $k$ steps. Indeed,  
among all the sets, those that exit the chain within $k$ steps will propagate freely forever 
either to  the left or to the right so that no further information is gained 
by increasing the resolution of these sets. Now, as we increase $k$ there are
more and more such sets that do not contribute to the entropy production
rate. This exponential decay continues for arbitrarily large $k$
because, in the steady state, there are arbitrarily fine variations in
the density of points on the chain, as can be seen from the singular nature of the SRB measure. 
\par The relation of this exponential decay to the escape rate is easy to 
derive. Indeed if, for an open system, the probability density decays like 
$\gamma$ (the escape rate), then one easily finds that the entropy should 
decay like $\gamma.$
This can be verified numerically. For the case of a system with mean drift
velocity $v,$ given in our case by Eq. (\ref{vdrift}), 
T\'el {\it et al.} \cite{vtb} showed that the escape rate formula of
Gaspard and Nicolis should be generalized to
\begin{equation}\label{esc}
\gamma = \fr{1}{4}\fr{v^2}{D} + \fr{\pi^2}{L^2}D.
\end{equation}
Fig. (11) shows a 
comparison between that formula and the numerically computed 
decay rate of the entropy production. The agreement is best around 
$s = 1/2.$ We believe that the small discrepancies for other values of $s$,
which are quadratic in $s$, may be due to small numerical errors and/or to 
next order corrections in $1/L.$
\subsection{Dissipative case}
We now switch to $\Phi_c$ and use formula \cfr{ent_prod_exp_c} to compute
the $0$-entropy production rate.
\begin{eqnarray}
\Delta_i S_0(n) &=&  \sum_{\omega_{0}}\mu_n(\omega_{0})
\ln\fr{\mu_n(\omega_{0})}{\nu^*(\omega_{0})\mu_n},\nonumber\\
&=& s\mu_{n - 1}\ln\fr{s\mu_{n - 1}}{(1 - s)\mu_n} + 
(1 - s)\mu_{n + 1}\ln\fr{(1 - s)\mu_{n + 1}}{s\mu_n},\nonumber\\
&=& s\mu_{n - 1}\ln\fr{\mu_{n - 1}}{\mu_n} + 
(1 - s)\mu_{n + 1}\ln\fr{\mu_{n + 1}}{\mu_n}\nonumber\\
&& + (s\mu_{n - 1} - (1 - s)\mu_{n + 1})\ln\fr{s}{1 - s},\label{tep}\\
&\equiv& \Delta_i S_0^{(vp)}(n) + \Delta_i S_0^{(d)}(n),\nonumber
\end{eqnarray}
where 
$$\Delta_i S_0^{(vp)}(n) = s\mu_{n - 1}\ln\fr{\mu_{n - 1}}{\mu_n} + 
(1 - s)\mu_{n + 1}\ln\fr{\mu_{n + 1}}{\mu_n}$$
is the contribution from the volume preserving part, identical to Eq. 
(\ref{0_ent_prod_vp}), and
\begin{equation}\label{0_ent_prod_dis}
\Delta_i S_0^{(d)}(n) = (s\mu_{n - 1} - (1 - s)\mu_{n + 1})\ln\fr{s}{1 - s}
= (2s - 1)(B - A\alpha^n)\ln\fr{s}{1 - s}
\end{equation}
reflects the presence of phase space contraction. 
Fig. (12) shows a numerical computation of this last term. 
It is constant and positive everywhere except in the vicinity of the
boundaries, where it may become negative. It is remarkable, though, that the 
total $0$-entropy production rate, Eq. (\ref{tep}), is positive everywhere, 
as shown in Fig. (13). Also, notice that the contribution due to the
dissipative term is generally much larger than the first term which is the 
only contribution in the  volume preserving case. 
\par Note that the dissipative term,
$$(\mu_n - 2A\alpha^n)(2 s - 1)\ln\fr{s}{1 - s},$$
is nothing but the phase space contraction rate \cfr{ps_cr} multiplied by
$\mu_n - 2A\alpha^n.$
We have thus shown that
\begin{equation}\label{ep_dis}
\Delta_i S_0(n) = D\fr{(\nabla\mu_n)^2}{\mu_n} - (\mu_n - 2A\alpha^n)
(\lambda_+ + \lambda_-).
\end{equation}
In the infinite volume limit $B$ takes on the values $1$ or $L+2$,
depending upon $\alpha$, and one can check that the entropy production rate becomes independent
of the boundary conditions. Indeed, the squared gradient term is overwhelmed 
by the second term, which is equal to the phase space contraction rate one
gets for periodic boundary conditions, and the third term, involving 
$A\alpha^n,$ is negligible in this limit. 

\par We also note that the thermostated system approach has been
applied by Chernov {\it et al.} \cite{cels} to systems in which no density 
gradient
is present. They consider the diffusion of a charged, moving particle in a
fixed array of hard scatterers, the periodic Lorentz gas, and use an
[electric field + thermostat] to generate an isokinetic electric current
in the system. In this case, there is no density gradient and all of
the irreversible entropy production comes from the phase space
contraction. The entropy production is then determined only by
the Lyapunov exponents and for small electric fields, at least, the
entropy production is proportional to the square of the electric field
with a coefficient that agrees with the predictions of irreversible
thermodynamics. 
\par Returning to our model, we can make use of the recursion relation, 
Eq. (\ref{ent_prod_rec}),
to compute the entropy production for different values of $k$. Although
the $k$ dependence will remain in the volume preserving part, we find
that the dissipative term does not depend on $k$ so that in
this case the largest contribution to the entropy production rate does
not depend on the resolution parameter. 
We remark that this property is specific to piecewise linear maps 
and should not be expected to be a general feature. 
Indeed, whereas, for a piecewise linear map, Eq.(\ref{ec4}) has a 
constant Jacobian in every single region of the partition, it will not
be so for non piecewise linear maps. As suggested by Eq. (\ref{ec6}), 
we will in general need to take the limit of infinite resolution to retrieve
the phase space contraction rate.

\section{Discussion}

\par We have shown that Gaspard's method \cite{gas1} of 
defining a coarse
grained Gibbs entropy and its rate of change for an Anosov-like, 
volume preserving dynamical
system can be generalized and extended to include non-volume
preserving Anosov-like systems which develop a nonequilibrium stationary
state SRB measure on an attractor. The rate of entropy production in
such a system is positive, and the total rate of entropy production in
a closed system is given by the negative of the sum of the Lyapunov
exponents of the map. 
Very close results have previously been obtained by Vollmer, T\'el and
Breymann, \cite{vtb}. Our contribution is mainly to show that 
Gaspard's coarse graining
method is a natural one to use in this context, and that it reveals
quite clearly the relation between the rate of irreversible entropy
production and the loss of information about the system's trajectory
due to the coarse graining. This method allows one to take a limit
where the coarse graining size is taken to zero after the
non-equilibrium steady state is reached, and in this limit we
recover the formula used by Ruelle \cite{rue1} to prove that the rate of 
entropy production is positive in the type of systems treated here.

\par The biased random walk models discussed here have a number of
interesting features. They are relatively simple to analyze, and they
exhibit an exponential growth and subsequent decay of the entropy production 
as the size of the coarse graining regions becomes smaller. However, for all 
values of $s$, except $s=1/2$, the density profile in the
non-equilibrium steady state is very unphysical, and the
large-system limit yields a trivial result where the density profile
is uniform except very close to the boundaries. The exponential
divergence of the rate of entropy production as the coarse graining
size gets large, at least over a range of $k$, is a striking difference 
between the multi-baker
chains we consider in this paper and that studied by Gaspard \cite{gas1}.
Indeed, whereas Gaspard showed, for $s=1/2$, that 
\begin{equation}\label{gaspard}
\lim_{k\rightarrow\infty}\lim_{(\nabla \mu_n)/\mu_n\rightarrow 0}
\lim_{L \rightarrow\infty}\fr{\mu_n}{(\nabla \mu_n)^2}\Delta_i S_k = D,
\end{equation}
we find, for $s\neq1/2$, that
\begin{equation}
\lim_{k\rightarrow\infty}\lim_{(\nabla \mu_n)/\mu_n\rightarrow 0}
\lim_{L \rightarrow\infty}\fr{\mu_n}{(\nabla \mu_n)^2}\Delta_i S^{(vp)}_k = 
D\left(\fr{(1 - s)^3 + s^3}{s(1 - s)}\right)^k.
\end{equation}

We believe that this point should be seen as a defect of the models we treat 
and will be addressed with more realistic models in further papers. 

T\'el, Vollmer and Breymann\cite{btv,vtb,vtbpre} consider a multi-baker 
model where the squares are organized in subsets that move to the right, left 
or stay within the squares. They show that there is a good scaling limit for
which the density profile is governed by 
a well defined Fokker Planck equation and is linear. 

We have chosen an alternative approach to the problem of finding
a realistic but analytically tractable model of a system with a
thermostat. In a subsequent paper\cite{gfd} we will discuss a model which can
be described as a random walk
driven by a thermostated external electric field and which takes the form
of a nonlinear baker map. Such
models mimic thermostated Lorentz gases which have been the subject of
a number of theoretical and computational studies\cite{cels,bec,romo,modero}. 
There we will also find an interesting transition of the dynamics from 
hyperbolic to non-hyperbolic behavior as a function of the strength of the 
external field, with dramatic consequences for the diffusive
properties of the system.

It has been realized for many years that the resolution of the ``Gibbs
paradox'' in entropy production, for many-particle systems, depends
upon using reduced distribution functions which themselves give a very
coarse grained description of a many particle system 
\cite{dorfman,tolman,ehrenfest}. Then the
macroscopic entropy production can be clearly identified, as in
Boltzmann's H-theorem, with the loss of information about the system's fine
grained phase space distribution with time. We have discussed here
a closely related and no less important mechanism for information loss and 
entropy production, namely the   
formation of fractal phase space structures in non-equilibrium
stationary states as the support for singular measures. The two mechanisms are related by the fact that
even for Hamiltionian systems, the support of the fine
grained, Gibbs distribution can evolve to a fractal that looks smooth in
phase space in some directions but highly singular in others. These
fractal structures require the application of coarse graining methods to correctly
describe irreversible processes in fluid systems and for an
understanding of why the reduced distribution functions approach their
equilibrium values in the course of time.

ACKNOWLEDGEMENTS: This paper is dedicated to E. G. D. Cohen in
celebration of his 75-th birthday - {\it ad mea v'esrim}. The authors wish 
to thank P. Gaspard, T. T\'el, J. Vollmer, W. Breymann, C. Dettmann, A. Latz 
and E. G. D. Cohen for a critical
reading of a draft of this paper which the authors feel has greatly
improved its clarity. They also thank S. Tasaki, H. van Beijeren, R. Klages,
and C. Ferguson for helpful discussions, as well as the referees for their
helpful remarks. J. R. D. wishes to acknowledge
support from the National Science Foundation under grant PHY~-96~-00428.

\begin{appendix}

\section{Properties of the SRB Measures}

In this appendix we briefly summarize the properties of the SRB measures for
the multibaker chains discussed in this paper and derive an analytical 
expression for the $k$-entropy, Eq. (\ref{ent_prod_exp_c}). For more details 
we refer to the papers of Gaspard \cite{gas1}, of Tasaki and Gaspard 
\cite{tasgas}, and of Tasaki, Gilbert, and Dorfman \cite{tgd}. We consider 
here the case of the volume contracting multibaker
map, $\Phi_{c}$, with flux boundary conditions. 
\par The SRB measure for
this system is obtained by using the Frobenius-Perron equation to
derive an expression for the cumulative measure in each unit square.
The Frobenius-Perron equation for the (singular) invariant density 
associated with this map is
\begin{equation}
\rho(\Gamma) =\int d\Gamma' \delta(\Gamma -\Phi_{c}(\Gamma'))\rho(\Gamma'),
\label{a1}
\end{equation}
or
\begin{equation}\label{a2}
\begin{array}{lcl@{\quad}r}
\rho(n,x,y) & = & \fr{s}{1 - s}\rho\left(n-1,sx,\fr{y}{1 - s}\right), &
0\leq y <1-s,\\
 & = & \fr{1 - s}{s}\rho\left(n + 1, s + (1 - s)x,
\fr{y - 1 + s}{s}\right), & 1-s \leq y \leq 1.
\end{array}
\end{equation}
The cumulative measure, $G(n,x,y)$ in each square is defined by
\begin{equation}
G(n, x, y) = \int_{0}^{x} dx'\int_{0}^{y} dy' \rho(n,x',y').
\label{a3}
\end{equation}
\par The measure of any region in a unit square can then be defined as the
difference of two cumulative functions. For example, the region
defined by a particular sequence $\{\omega_{0},...,\omega_{k-1}\}$
in square $n$ is a horizontal strip extending over the full $x$
interval, and contained between $y(\omega_{0},...,\omega_{k-1})$
and $y(\omega_{0},...,\omega_{k-1} + 1)$, so that
\begin{equation}
\mu_{n}(\omega_0,...,\omega_{k-1}) =
G(n,1,y(\omega_0,...,\omega_{k-1} + 1)) -
G(n,1,y(\omega_{0},...,\omega_{k-1})),
\label{a4}
\end{equation}
where we have introduced the notation
\begin{equation}\label{omega+1}
\begin{array}{lcl@{\quad}r}
\omega_{j - 1}, \omega_j + 1 &=& \omega_{j - 1}, 1, 
&\omega_j = 0,\\
&=& \omega_{j - 1} + 1, 0, &\omega_j = 1,
\end{array}
\end{equation}
with the convention that $y(1,\ldots, 1, 1 + 1) = 1.$ 
\par We can also write explicitely
\begin{equation}
y(\omega_{0},...,\omega_{k-1}) = \omega_0(1 - s) + \sum_{j = 1}^{k - 1}
\omega_j (1 - s) \nu^*(\omega_0, \ldots,\omega_{j - 1}),
\end{equation}
where $\nu^*$ is as defined by Eq. (\ref{nu*}). In particular, we note that
\begin{eqnarray}
y(0, \omega_{1},...,\omega_{k-1}) &=& 
(1 - s)y(\omega_{1},...,\omega_{k-1}),\label{y(0...)}\\
y(1, \omega_{1},...,\omega_{k-1}) &=& 
1 - s(1 - y(\omega_{1},...,\omega_{k-1})).\label{y(1...)}
\end{eqnarray}
These equations will be used in the sequel.
\par It follows from the Frobenius-Perron equation (\ref{a2}) that $G(n,x,y)$
satisfies the equation
\begin{equation}\label{a5}
\begin{array}{lcl@{\quad}r}
G(n,x,y) & = & G\left(n - 1, sx, \fr{y}{1 - s}\right), &0 \leq y <1-s,
\\
& = & G(n - 1, sx, 1)+G\left(n + 1, (1 - s)x + s, \fr{y - 1 + s}{s}\right) & \\
& & - G\left(n + 1, s, \fr{y - 1 + s}{s}\right), & 1 - s \leq y \leq 1.
\end{array}
\end{equation}
Then, the total measure of the $n$-th square, $\mu_{n}$, satisfies the equation
\begin{equation}
\mu_{n}=G(n, 1, 1) = G(n - 1, s, 1)+G(n + 1, 1, 1)-G(n + 1, s, 1).
\label{a6}
\end{equation} 
It is now possible to see that there is a solution of Eq. (\ref{a5})
for $G(n,x,y)$ which has the form
\begin{equation}\label{a7}
G(n, x, y)=x [y(\mu_{n} - B) + B T_{n}(y)],
\end{equation}
where $\mu_n$ is a solution of Eq. (\ref{ss}), $B$ is given by 
Eq. (\ref{fbc_GMB}), and $T_{n}(y)$ satisfies the recursion relation
\begin{equation}\label{a8}
\begin{array}{lcl@{\quad}r}
T_{n}(y) & =& s T_{n - 1}\left(\fr{y}{1 - s}\right), &0\leq y < 1 - s, \\
& =& s + (1 - s)T_{n + 1}\left(1 - \fr{1 - y}{s}\right), & 1 - s\leq y\leq 1.
\end{array}
\end{equation}
The boundary conditions on the functions $T_{n}(y)$ are 
$T_{0}(y)=T_{L + 1}(y) = y$. These functions will be referred to as 
incomplete, see Tasaki and Gaspard \cite{tasgas}, as opposed to the function
$T(y)$ which appears in the case of periodic boundary conditions, see 
Tasaki {\it et al.} \cite{tgd},
\begin{equation}\label{lim_T}
\begin{array}{lcl@{\quad}r}
T(y) & =& s T\left(\fr{y}{1 - s}\right), & 0\leq y < 1 - s, \\
& =& s + (1 - s)T\left(1 - \fr{1 - y}{s}\right), & 1 - s\leq y\leq 1.
\end{array}
\end{equation}

\par The solution given by Eq. (\ref{a7}) leads immediately to the
recursion relation, Eq. (\ref{ss}), for the measures
$\mu_{n}$. Furthermore, this expression for $G(n,x,y)$ has the form
expected for the cumulative distribution of an SRB measure~: it is
smooth (actually uniform) in the expanding direction, and singular in the
contracting direction. The singularity in the
contracting, or $y$-direction, can be seen when the recursion relations
are solved for the functions $T_{n}(y)$. In the limit where the boundaries 
are infinitely far away, $T_n(y)$ is replaced by the limiting function
$T(y)$, which is a continuous function that has zero derivatives almost
everywhere. It is the singularity of $T(y)$ which prevents the measure
from having a well behaved density, and requires the use of the coarse
graining procedure described in the body of this paper. Fig. (14) shows, 
for $s = .55$, the limiting function $T(y) - y$, singular on every length 
scale. Figs. (15-20) show, for the same value of the parameter $s,$ the 
incomplete functions $T_n(y) - y$ truncated due to the boundary conditions 
$T_0(y) = T_{L + 1}(y) = y.$ For some fixed length scale, these 
functions quickly become singular (their derivatives have discontinuities 
inside the corresponding $y$ interval) as we move away from the boundaries. 
Entropy production is positive in the $y$ intervals where $T_n(y)$ has 
singularities.  
\par Before we proceed to the derivation of the $k$-entropy, we make use
of Eqs. (\ref{y(0...)}, \ref{y(1...)}) to derive a useful expression of the 
$T_n(y).$ Note that
\begin{eqnarray*}
T_n\Big(y(0, \omega_{1},...,\omega_{k-1})\Big) &=& 
T_n\Big((1 - s)y(\omega_{1},...,\omega_{k-1})\Big),\\
&=& s T_{n - 1}\Big(y(\omega_{1},...,\omega_{k-1})\Big),\\
T_n\Big(y(1, \omega_{1},...,\omega_{k-1})\Big) &=& 
T_n\Big(1 - s(1 - y(\omega_{1},...,\omega_{k-1}))\Big),\\ 
&=& s + (1 - s)T_{n + 1}\Big(y(\omega_{1},...,\omega_{k-1})\Big).
\end{eqnarray*}
As long as we are far enough from the boundaries, we can replace
the $T_n(y)$ by their limiting values $T(y)$ (this amounts to assuming 
$1 << n << L$). In this case, it follows that
\begin{equation}\label{T_exp}
T(y(\omega_{0},...,\omega_{k-1})) = s\omega_0 + 
\sum_{j = 1}^{k - 1} s\omega_j\nu(\omega_0, \ldots, \omega_{j - 1}),
\end{equation}
where $\nu$ is as defined by Eq. (\ref{nu}).
\par With the help of Eq. (\ref{a4}), we can make use of Eq. (\ref{a7}) to 
rewrite the entropy production rate formula, Eq. (\ref{ent_prod_exp_c}). 
Ignoring the boundary effects, Eq. (\ref{a4}) gives
\begin{eqnarray*}
\mu_n(\omega_0, \ldots, \omega_{k - 1}, 0) &=&
B\left[T(y(\omega_0, \ldots, \omega_{k - 1}, 1)) - 
T(y(\omega_0, \ldots, \omega_{k - 1}))\right] \\
&&+ A\alpha^n(1 - s) \nu^*(\omega_0, \ldots, \omega_{k - 1}),\\
\mu_n(\omega_0, \ldots, \omega_{k - 1}, 1) &=&
B\left[T(y(\omega_0, \ldots, \omega_{k - 1} + 1)) - 
T(y(\omega_0, \ldots, \omega_{k - 1}, 1))\right] \\
&&+ A\alpha^n s \nu^*(\omega_0, \ldots, \omega_{k - 1}).\\
\end{eqnarray*}
Writing
\begin{eqnarray}
&&\begin{array}{lcl}
\Delta T_a &=& T(y(\omega_0, \ldots, \omega_{k - 1}, 1)) - 
T(y(\omega_0, \ldots, \omega_{k - 1})),\\
\Delta T_b &=& T(y(\omega_0, \ldots, \omega_{k - 1} + 1)) - 
T(y(\omega_0, \ldots, \omega_{k - 1}, 1)),
\end{array}\label{DT}\\
&&\varepsilon = A\alpha^n\nu^*(\omega_0, \ldots, \omega_{k - 1}),\label{eps}
\end{eqnarray}
the $k$-entropy production rate becomes
\begin{eqnarray}
\Delta_i S_k(n) = \sum_{\omega_0, \ldots, \omega_{k - 1}}&&\left[
(B\Delta T_a + (1 - s)\varepsilon)\ln\fr{B\Delta T_a/(1 - s) + \varepsilon}
{B\Delta T_a + B\Delta T_b + \varepsilon}\right.\nonumber\\
&& + \left.
(B\Delta T_b + s\varepsilon)\ln\fr{B\Delta T_b/s + \varepsilon}
{B\Delta T_a + B\Delta T_b + \varepsilon}\right].\label{DiSk1}
\end{eqnarray}
In the limit $1 << n << L,$ $A\alpha^n << 1$ so that we can expand the 
logarithms. Up to second order, we find
\begin{eqnarray}
\Delta_i S_k(n) = \sum_{\omega_0, \ldots, \omega_{k - 1}}&&\left[
(B\Delta T_a + (1 - s)\varepsilon)\ln\fr{\Delta T_a/(1 - s)}
{\Delta T_a + \Delta T_b}
+ (B\Delta T_b + s\varepsilon)\ln\fr{B\Delta T_b/s}
{\Delta T_a + \Delta T_b}\right.\nonumber\\
&&\left. + \fr{\varepsilon^2}{2B}\left(\fr{(1 - s)^2}{\Delta T_a} + 
\fr{s^2}{\Delta T_b} - \fr{1}{\Delta T_a + \Delta T_b}\right)\right].
\label{DiSk2}
\end{eqnarray}
The following properties of $\Delta T_a$ and $\Delta T_b$ follow easily from 
Eqs. (\ref{T_exp}, \ref{DT})~:
\begin{eqnarray*}
\Delta T_a &=& s(\Delta T_a + \Delta T_b),\\
\Delta T_b &=& (1 - s)(\Delta T_a + \Delta T_b).
\end{eqnarray*}
Eq. (\ref{DiSk2}) thus becomes
\begin{equation}\label{DiSk3}
\Delta_i S_k(n) = \Delta_i S_k^{(vp)}(n) + \Delta_i S_k^{(d)}(n),
\end{equation}
where we have set
\begin{eqnarray}
\Delta_i S_k^{(vp)}(n) &=& \sum_{\omega_0, \ldots, \omega_{k - 1}}
\fr{\varepsilon^2}{2B(\Delta T_a + \Delta T_b)}
\left(\fr{(1 - s)^2}{s} + \fr{s^2}{1 - s} - 1\right),\nonumber\\
&=& \fr{(1 - 2s)^2}{2s(1 - s)}\fr{A^2\alpha^{2n}}{B}
\sum_{\omega_0, \ldots, \omega_{k - 1}} 
\fr{\nu^*(\omega_0, \ldots, \omega_{k - 1})^2}{\Delta T_a + \Delta T_b},
\label{DiSk_vp}
\end{eqnarray}
and
\begin{eqnarray}
\Delta_i S_k^{(d)}(n) &=& (2s - 1)\ln\fr{s}{1 - s}
\sum_{\omega_0, \ldots, \omega_{k - 1}}[B(\Delta T_a + \Delta T_b) - 
\varepsilon],\nonumber\\
&=&(2s - 1)\ln\fr{s}{1 - s}(B - A\alpha^n).\label{DiSk_d}
\end{eqnarray}
To derive Eq. (\ref{DiSk_d}) we made use of the property
$$\sum_{\omega_0, \ldots, \omega_{k - 1}}\Delta T_a + \Delta T_b = 1,$$
which follows from the identity
\begin{equation}\label{DT_nu}
\Delta T_a + \Delta T_b = \nu(\omega_0, \ldots, \omega_{k - 1}),
\end{equation}
that follows itself easily from Eqs. (\ref{T_exp}, \ref{DT}).
\par The comparison between Eqs. (\ref{DiSk3}-\ref{DiSk_d}) and
the corresponding 0-entropy production rate, Eqs. (\ref{ep_vp},
\ref{tep}, \ref{0_ent_prod_dis}) is straightforward. Set $k = 0$ in 
Eq. (\ref{DiSk_vp}). Then there is no index over which to sum, $\nu^*$ is 
replaced by 1, and
$$\Delta T_a + \Delta T_b = T(1) - T(0) = 1.$$
As of Eq. (\ref{DiSk_d}), it is just the same as Eq. (\ref{0_ent_prod_dis}),
thus confirming the $k$ independence of that part of the $k$-entropy 
production rate.
\par Let us now investigate the $k$ dependence of Eq. (\ref{DiSk_vp}). To
this effect, we rewrite Eq. (\ref{DiSk_vp}) using Eq. (\ref{DT_nu})~:
\begin{eqnarray}
\Delta_i S_k^{(vp)}(n) &=& \fr{(1 - 2s)^2}{2s(1 - s)}\fr{A^2\alpha^{2n}}{B}
\sum_{\omega_0, \ldots, \omega_{k - 1}}
\fr{\nu^*(\omega_0, \ldots, \omega_{k - 1})^2}
{\nu(\omega_0, \ldots, \omega_{k - 1})},\nonumber\\
&=& \fr{(1 - 2s)^2}{2s(1 - s)}\fr{A^2\alpha^{2n}}{B}
\left(\sum_{\omega}
\fr{\nu^*(\omega)^2}
{\nu(\omega)}\right)^k,\nonumber\\
&=& \fr{(1 - 2s)^2}{2s(1 - s)}\fr{A^2\alpha^{2n}}{B}
\left(\fr{(1 - s)^3 + s^3}{s(1 - s)}\right)^k.\label{DiSk4}
\end{eqnarray}
\par We thus conclude that the volume preserving part of the $k$-entropy 
diverges exponentially with $k$ ! This is quite remarkable as it differs
dramatically from the expression Gaspard \cite{gas1} gave for the case 
$s = .5$ for which the $k$ divergence is linear and is next order in the small 
parameter and thus vanishes in the limit (\ref{gaspard}). In our case, the
divergence is of the same order in $A\alpha^n,$ which illustrates the
breakdown of the $k$-entropy production for biased random walk models
considered here.  

\end{appendix}

\newpage
\centerline{\bf FIGURE CAPTIONS}

{\it Figure 1.} The volume preserving, deterministic version of the
biased random walk, $\Phi$, defined by Eq. (\ref{GMB}).

{\it Figure 2.} The dissipative, deterministic version of the biased
random walk, $\Phi_c$, defined by Eq. (\ref{DMB}).

{\it Figure 3.} The $(1, 1)$ and $(0, 2)$-sets of $\Phi$. The dots indicate
the separation between the indices $l$ and $k$.

{\it Figure 4.} The $(1, 1)$ and $(0, 2)$-sets of $\Phi_c$. The dots indicate
the separation between the indices $l$ and $k$.

{\it Figure 5.} The stationary state distribution $\mu_n$, solution of Eqs.
(\ref{ss_GMB}, \ref{fbc_GMB}), on a chain of 
$L = 100$ sites as a function of the lattice coordinate $n$ ($s = .5$ is 
the solid line, $s = .45$ the dadhed line and $s = .55$ the long dashed
line).

{\it Figure 6.} 0-entropy production, $\Delta_{i} S_0(n)$, in the 
volume preserving case, Eq. (\ref{0_ent_prod_vp}) for a chain of $L = 100$ 
sites as a function of the lattice coordinate $n$ and for 
$s = .1,\ldots,.9$ from left to right (the solid line corresponds to $s = .5$).

{\it Figure 7.} k-entropy production, $\Delta_{i} S_{k}(n)$, in the volume 
preserving case, for $k = 0,\ldots,9$, numerically computed using Eq. 
(\ref{ent_prod_rec}), $s = 0.45$ and $L=100$.

{\it Figure 8.} k-entropy production, $\Delta_{i} S_{k}(n)$, in the volume 
preserving case, for $k = 0,\ldots,9$, numerically computed using Eq. 
(\ref{ent_prod_rec}), $s = 0.55$ and $L=100$.

{\it Figure 9.} k-entropy production, $\Delta_{i} S_{k}(n = 50)$, in the 
volume preserving case, as a function of $k,$ numerically computed using Eq. 
(\ref{ent_prod_rec}), $L=100$. Both $s = 0.45$ and $s = 0.55$ are displayed.
On the long range scale, the entropy production decays exponentially as
a function of $k$ as the invariant measure gets mostly smooth on the
corresponding scales.

{\it Figure 10.} Blow up of Fig. (10) for $k = 0,\ldots, 50.$ The $k$-entropy
diverges exponentially with $k$. The slope is given by Eq. (\ref{DiSk4}).

{\it Figure 11.} A comparison between the escape rate given by Eq. (\ref{esc})
(solid line) and the numerically computed decay rate (diamonds) of the entropy 
production at large values of $k$ as a function of $s$ ($L=100$ and the decay 
rate was measured at $k=8000$).

{\it Figure 12.} Dissipative part, $\Delta_{i} S_0^{(d)}(n)$, of the 
0-entropy production, Eq. (\ref{0_ent_prod_dis}), for a chain of $L = 100$ 
sites as a function of the lattice coordinate $n$. $s = .1,\cdots,.9$.
The $s$-values increase from top to bottom.

{\it Figure 13.} Total 0-entropy production, Eq. (\ref{tep}), in the 
dissipative case for a chain of $L = 100$ sites as a function of the 
lattice coordinate $n$. $s = .1,\cdots,.9$.
The $s$-values increase from top to bottom.

{\it Figure 14.} The limiting function $T(y) - y$, Eq. (\ref{lim_T}), 
$s = .55$.

{\it Figures 15-20.} The incomplete functions $T_n(y) - y,$ Eq. (\ref{a8}), 
$n = 1, 3, 5$ and $n = 96, 98, 100$ with the boundary conditions 
$T_0(y) = T_{101}(y) = y$ and $s = .55$.

\newpage

\begin{figure}[htb]
\centerline{\psfig{figure=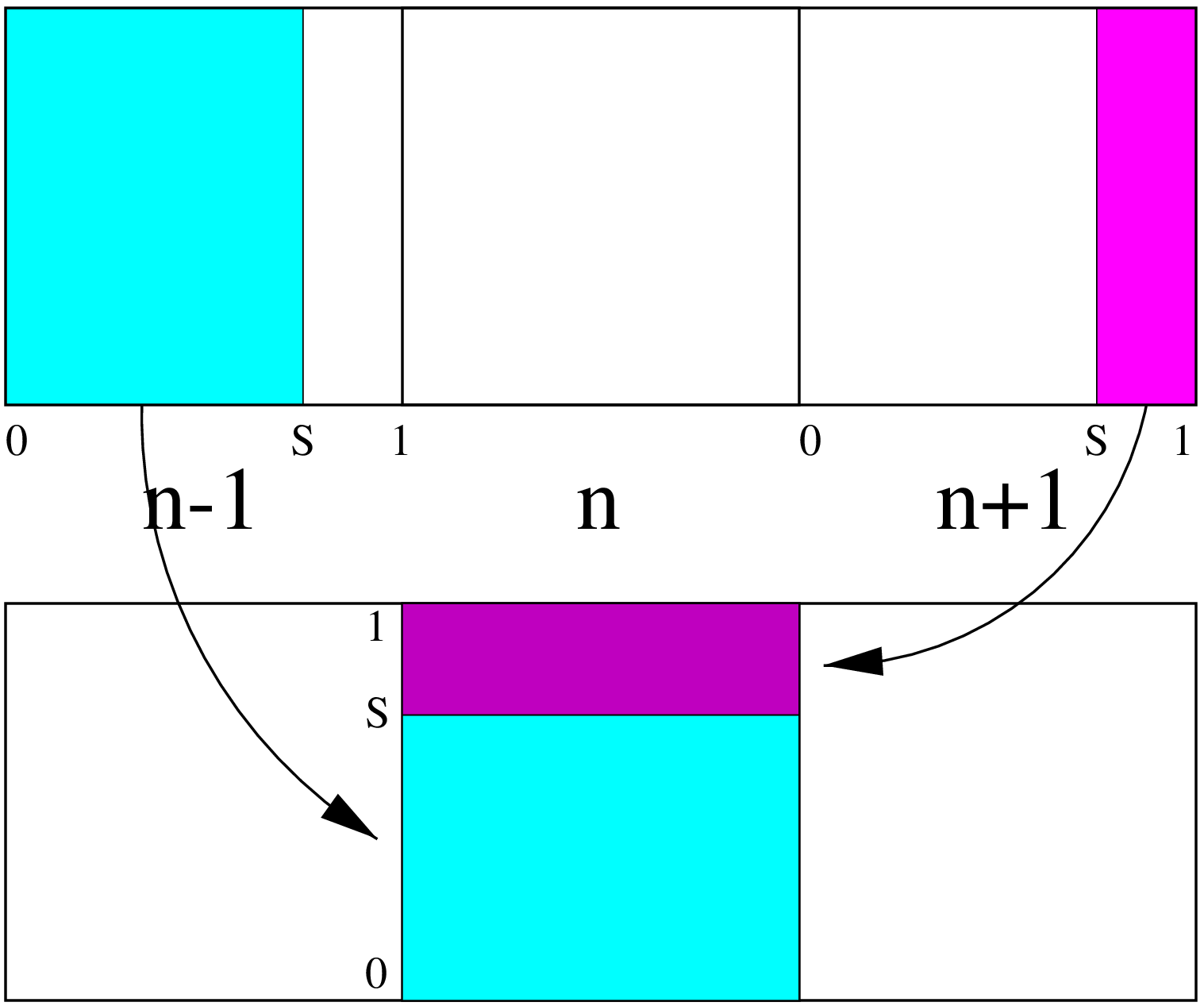}}
\caption{}
\end{figure}

\newpage

\begin{figure}[htb]
\centerline{\psfig{figure=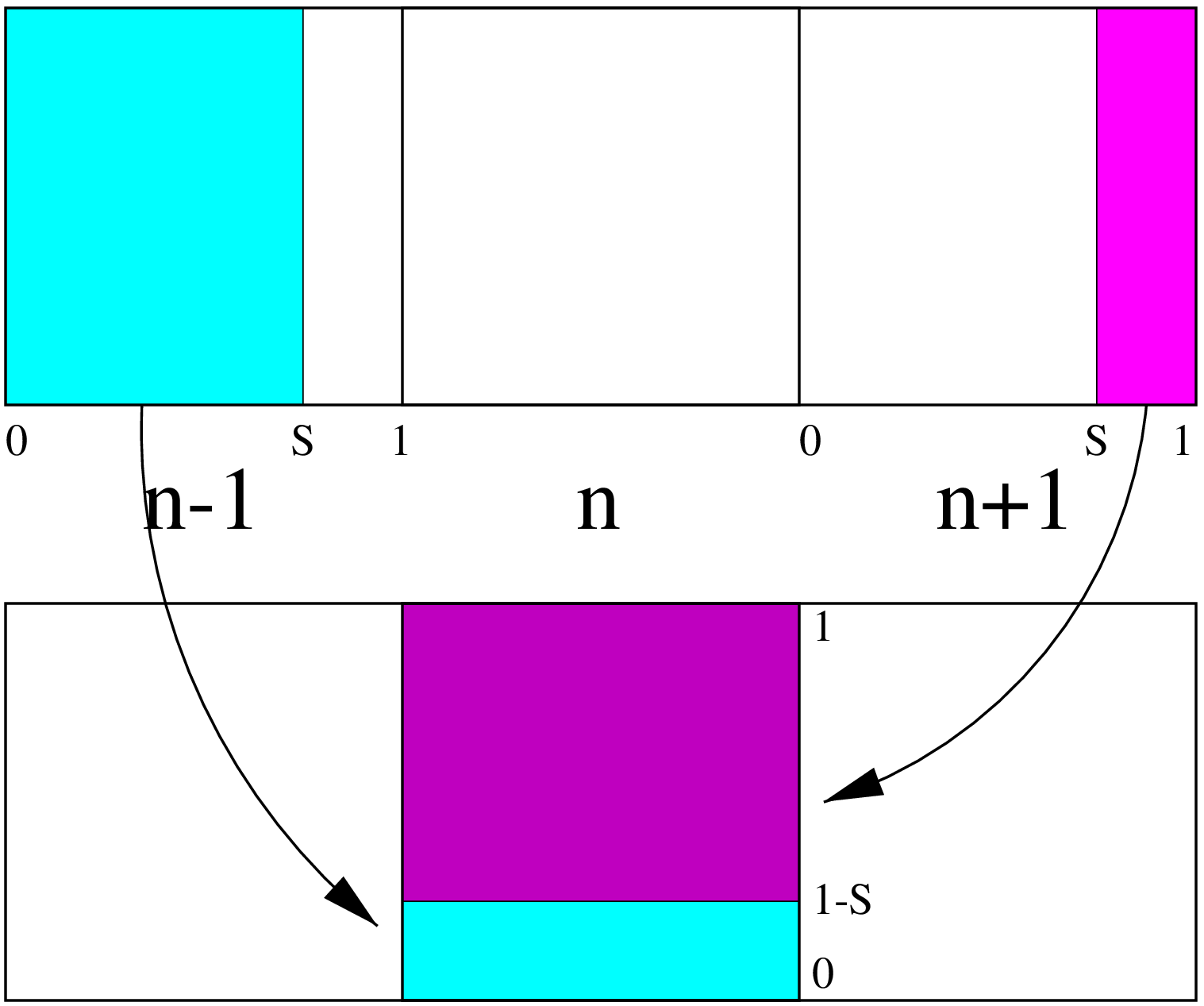}}
\caption{}
\end{figure}

\newpage

\begin{figure}[htb]
\centerline{\psfig{figure=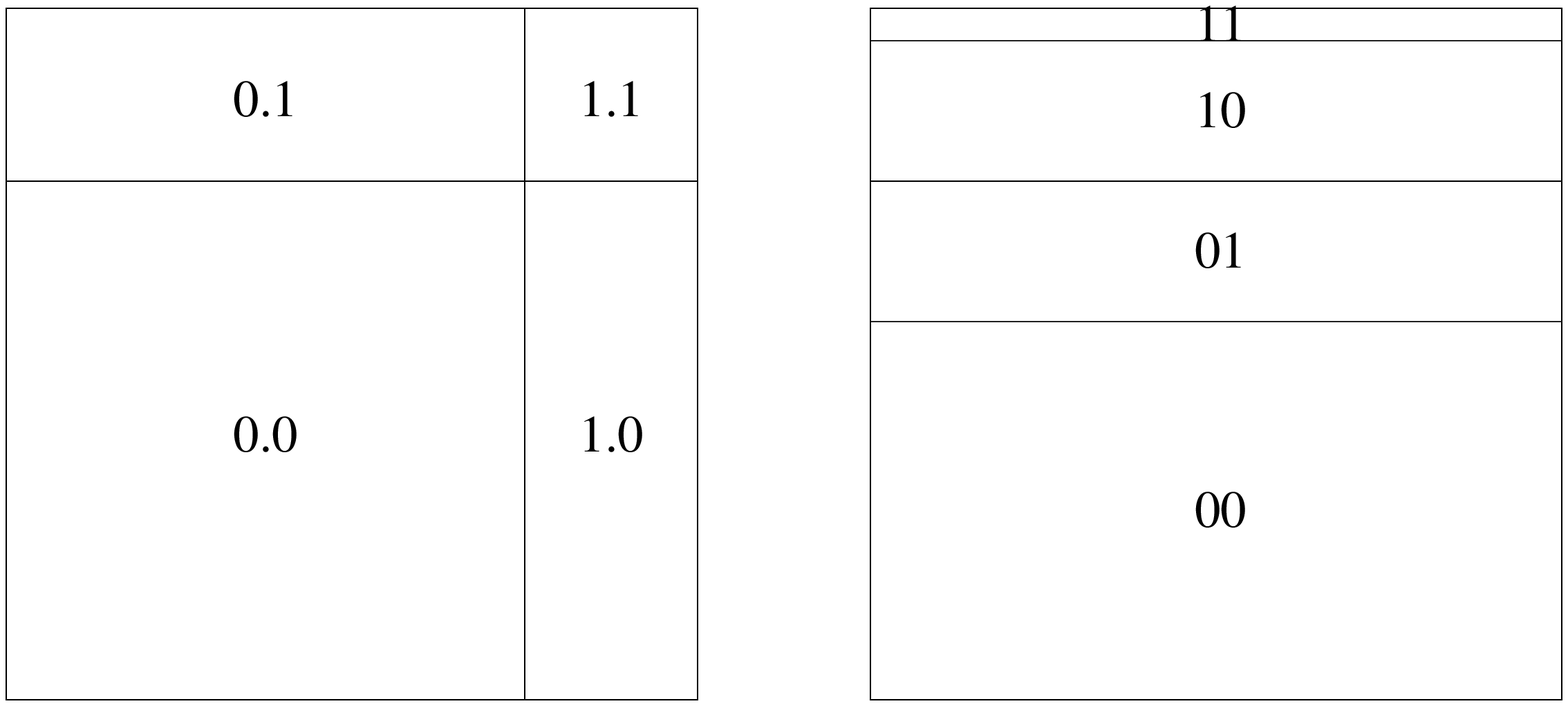,width=16cm}}
\caption{}
\end{figure}

\newpage

\begin{figure}[htb]
\centerline{\psfig{figure=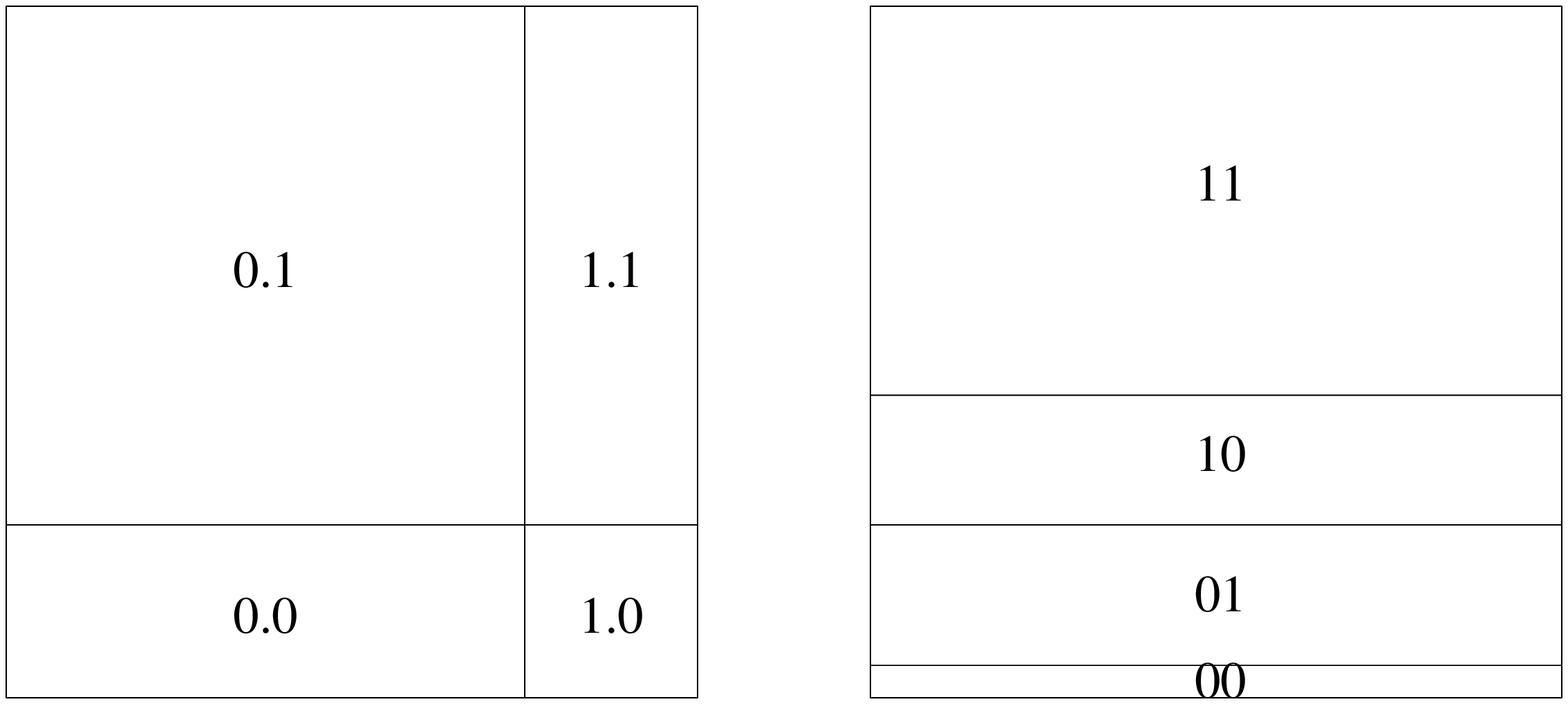,width=16cm}}
\caption{}
\end{figure}

\newpage

\begin{figure}[htb]
\centerline{\psfig{figure=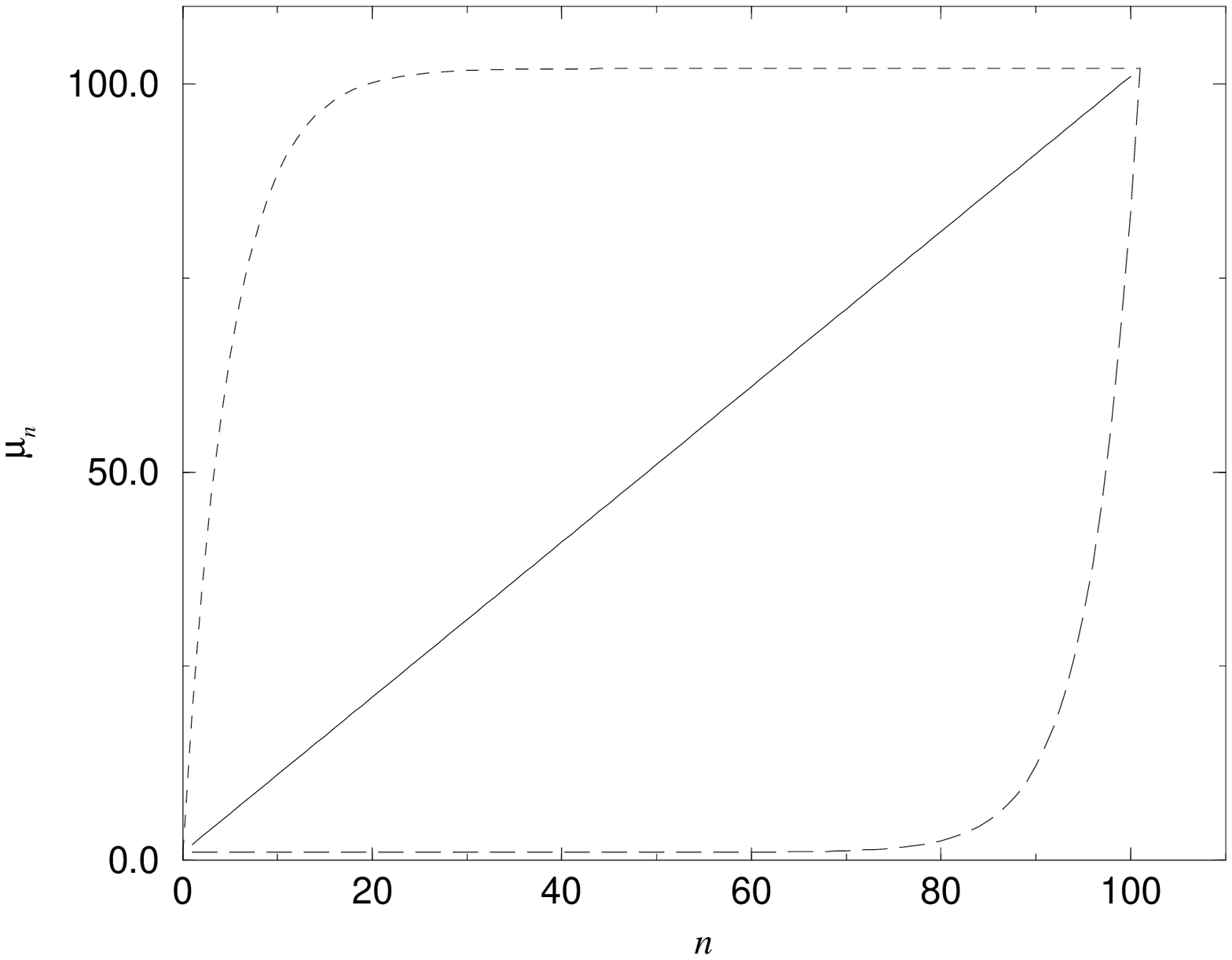}}
\caption{}
\end{figure}

\newpage

\begin{figure}[htb]
\centerline{\psfig{figure=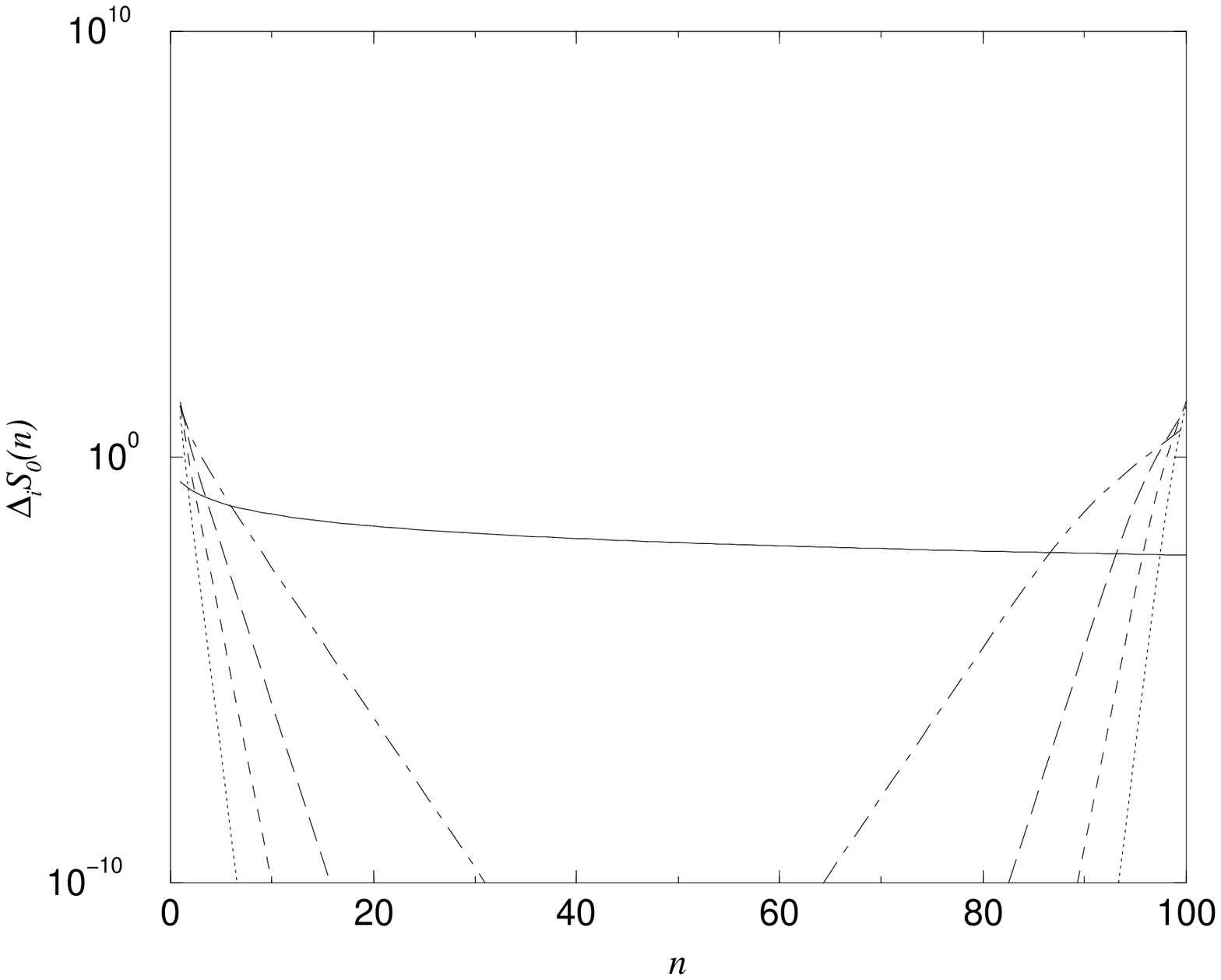}}
\caption{}
\end{figure}

\newpage

\begin{figure}[htb]
\centerline{\psfig{figure=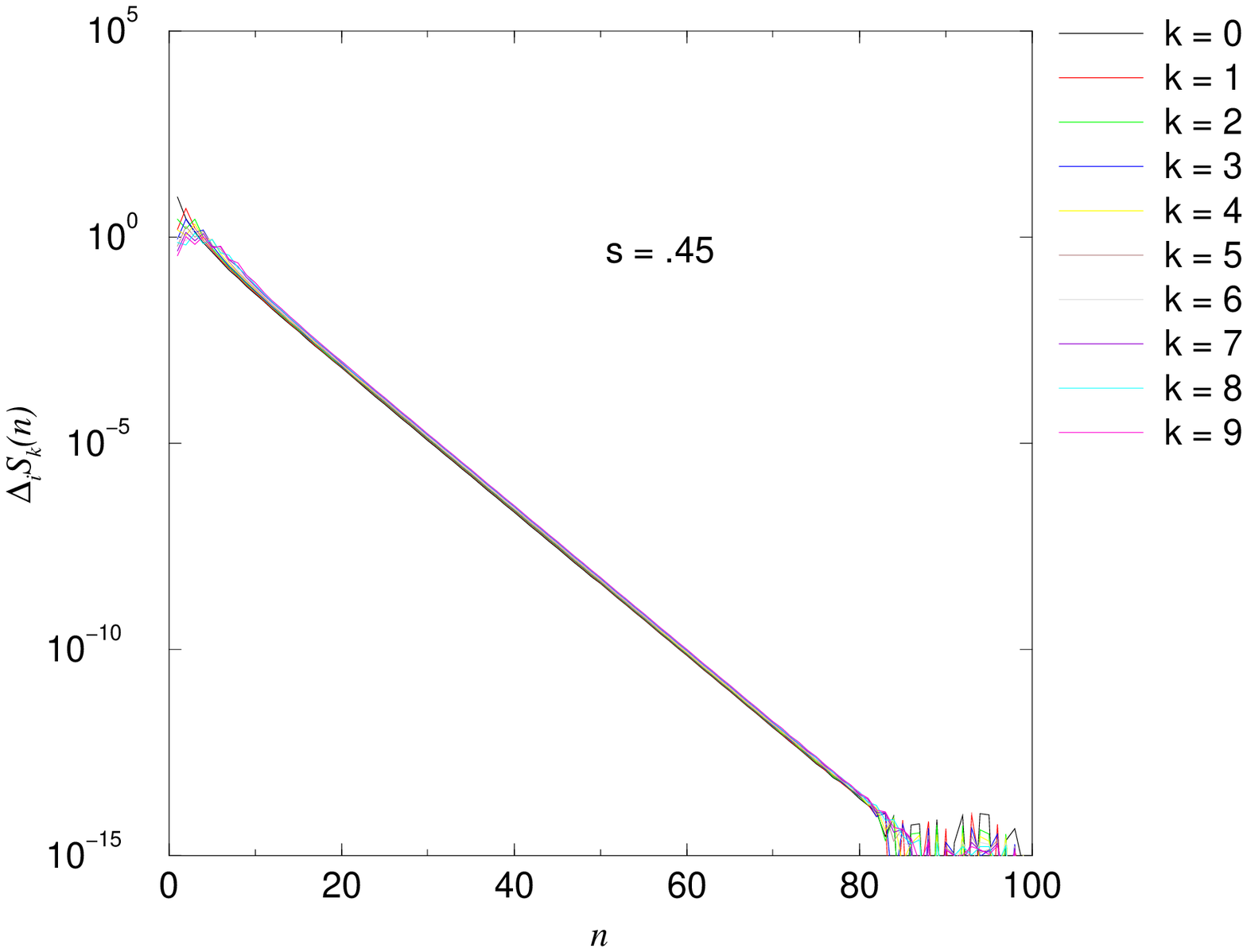}}
\caption{}
\end{figure}

\newpage

\begin{figure}[htb]
\centerline{\psfig{figure=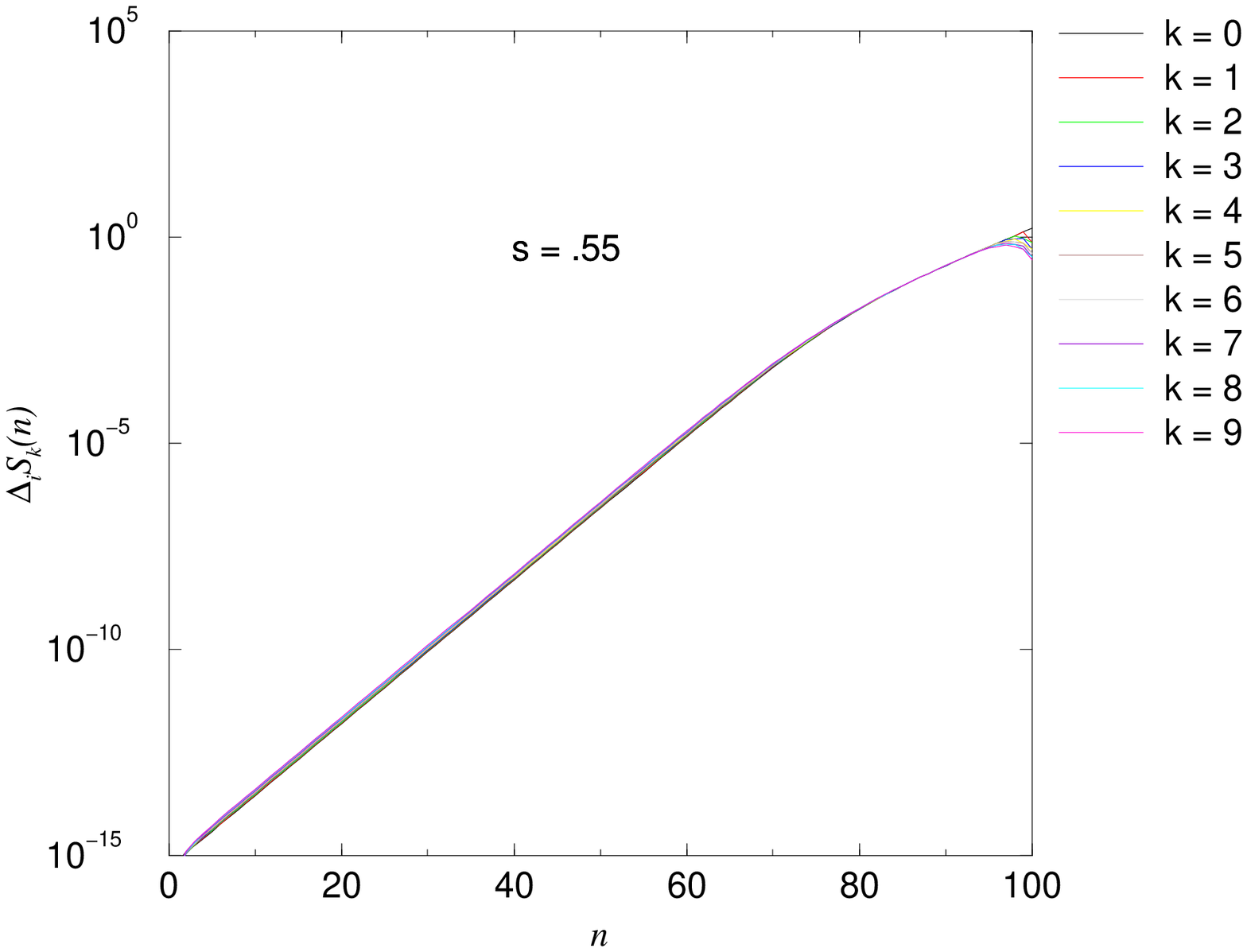}}
\caption{}
\end{figure}

\newpage

\begin{figure}[htb]
\centerline{\psfig{figure=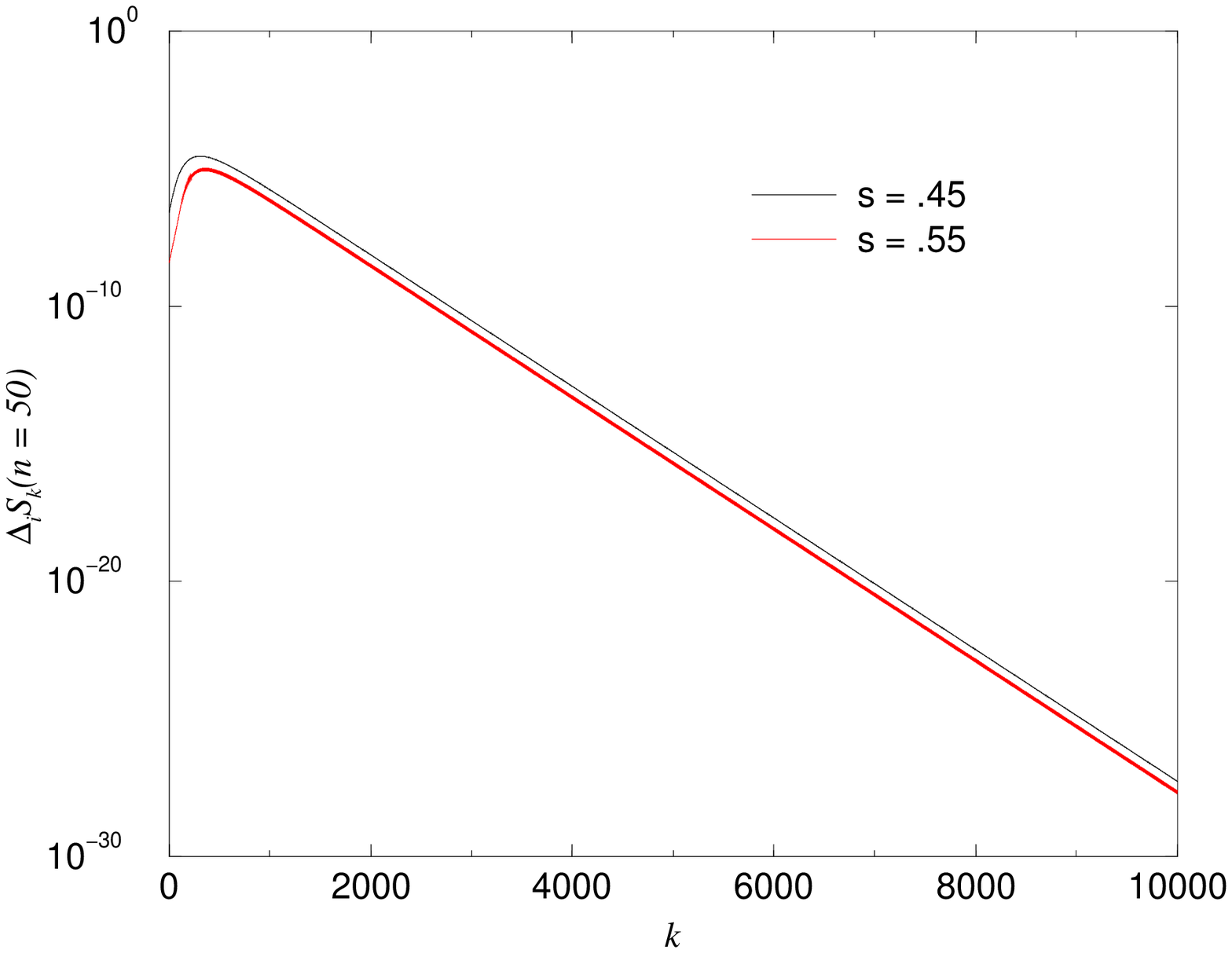}}
\caption{}
\end{figure}

\newpage

\begin{figure}[htb]
\centerline{\psfig{figure=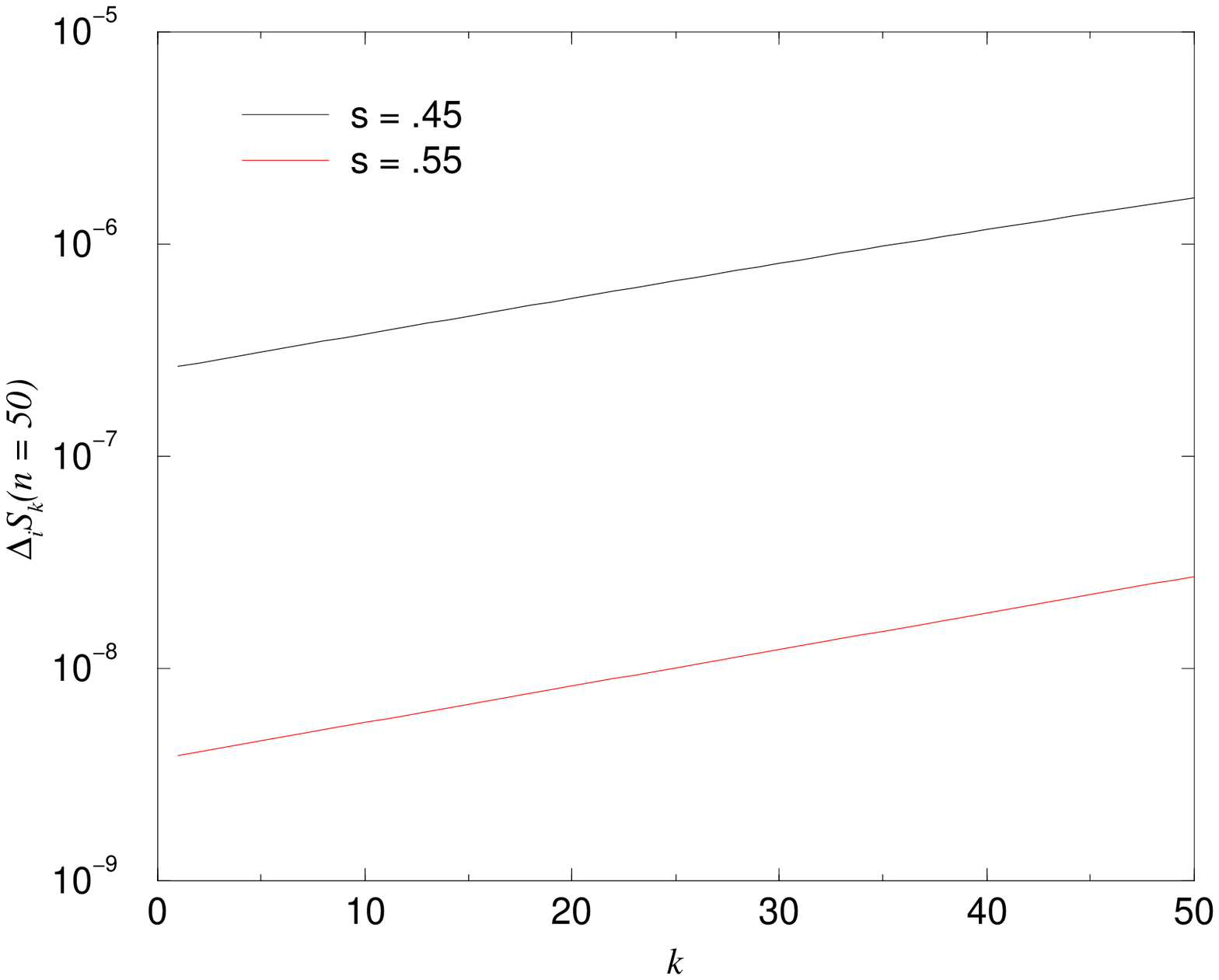}}
\caption{}
\end{figure}

\newpage

\begin{figure}[htb]
\centerline{\psfig{figure=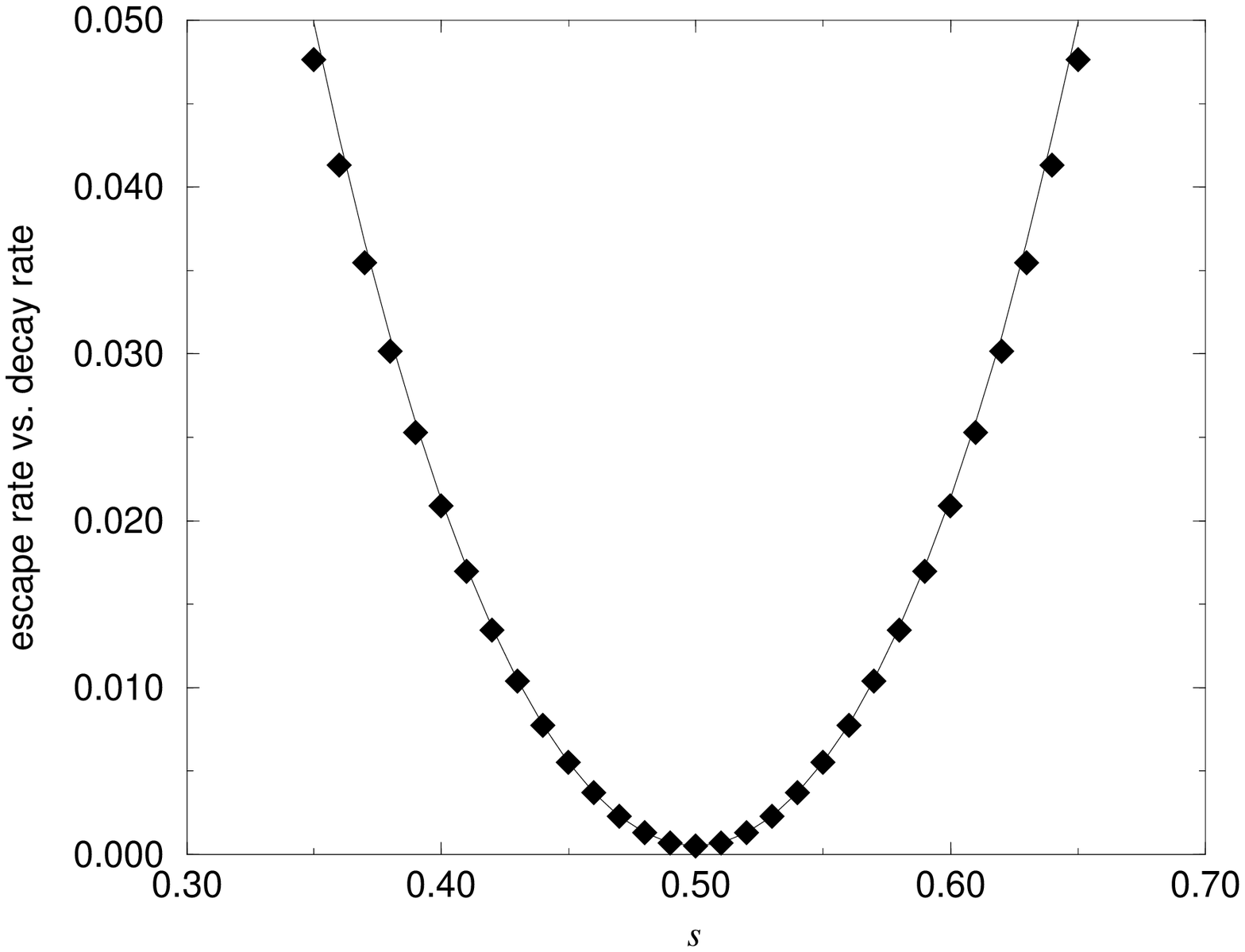}}
\caption{}
\end{figure}

\newpage

\begin{figure}[htb]
\centerline{\psfig{figure=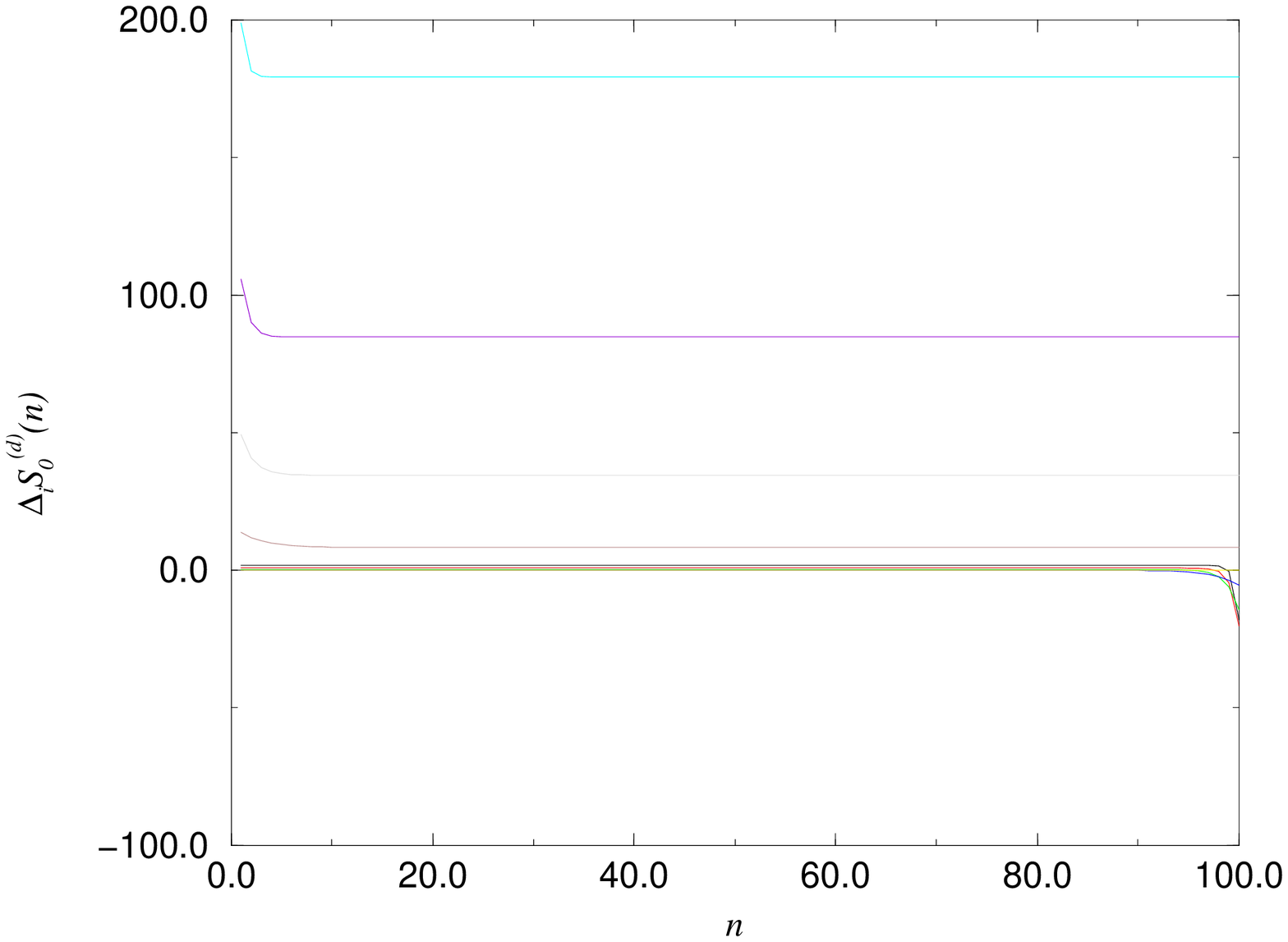}}
\caption{}
\end{figure}

\newpage

\begin{figure}[htb]
\centerline{\psfig{figure=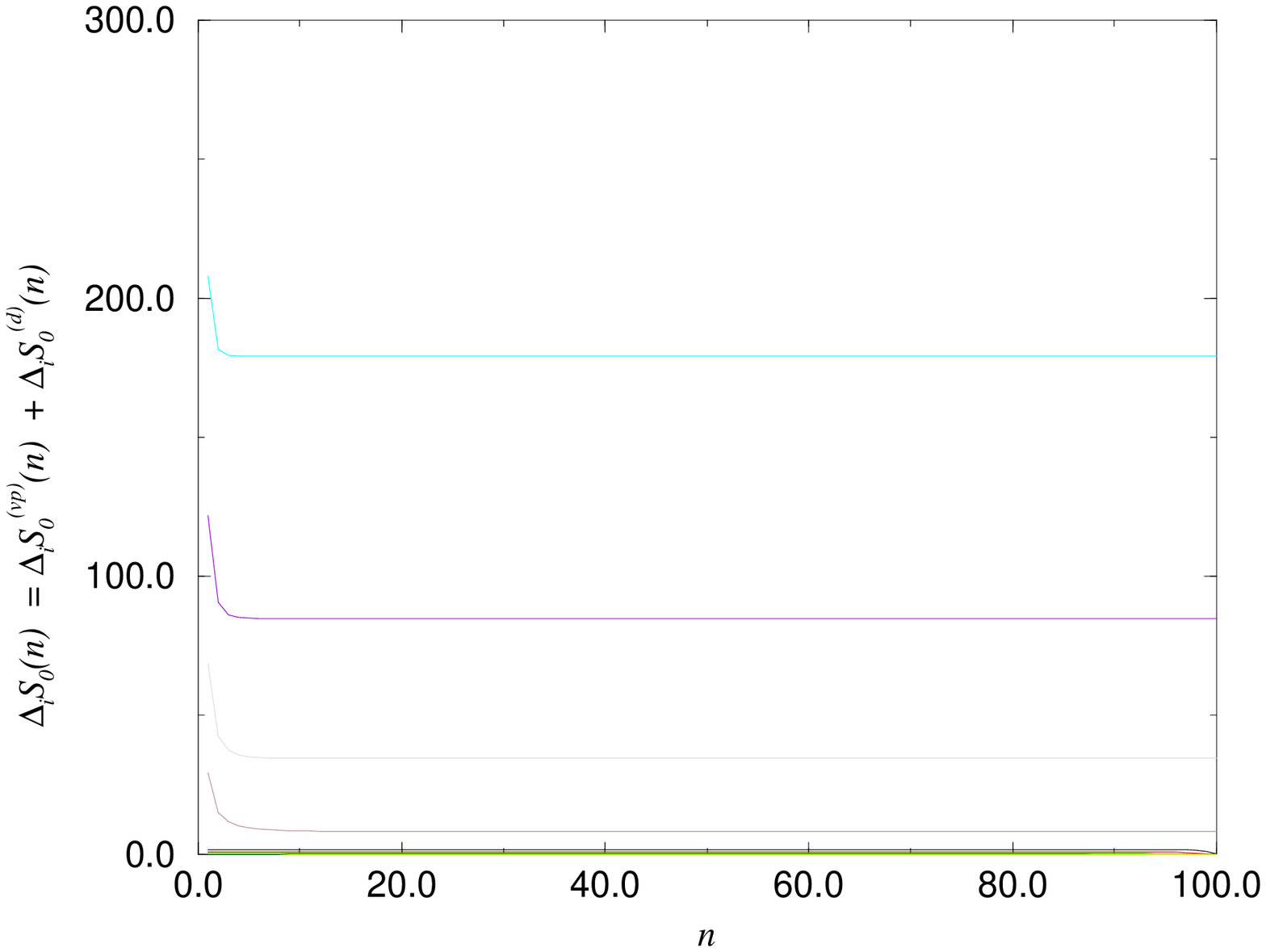}}
\caption{}
\end{figure}

\newpage

\begin{figure}[htb]
\centerline{\psfig{figure=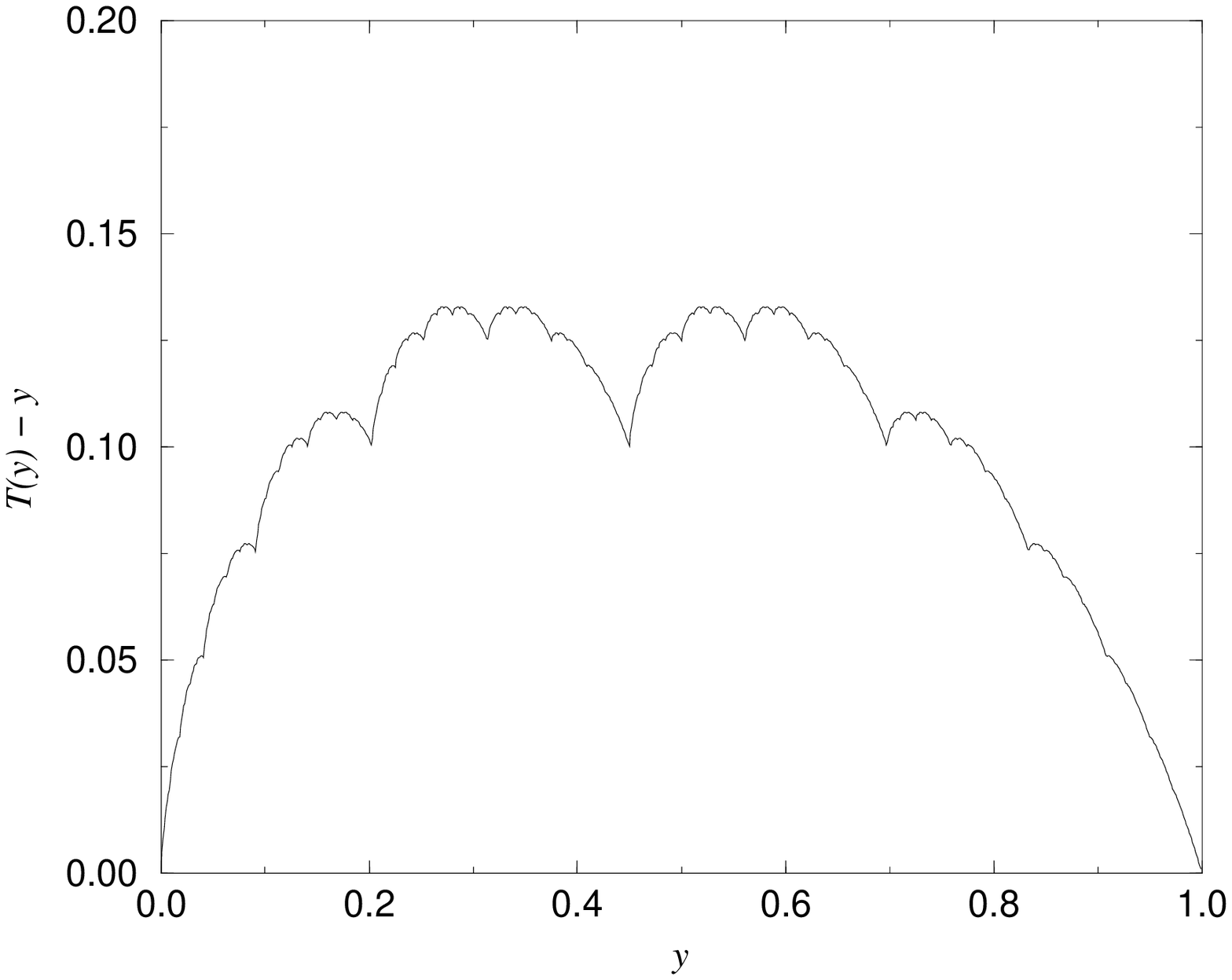}}
\caption{}
\end{figure}

\newpage

\begin{figure}[htb]
\centerline{\psfig{figure=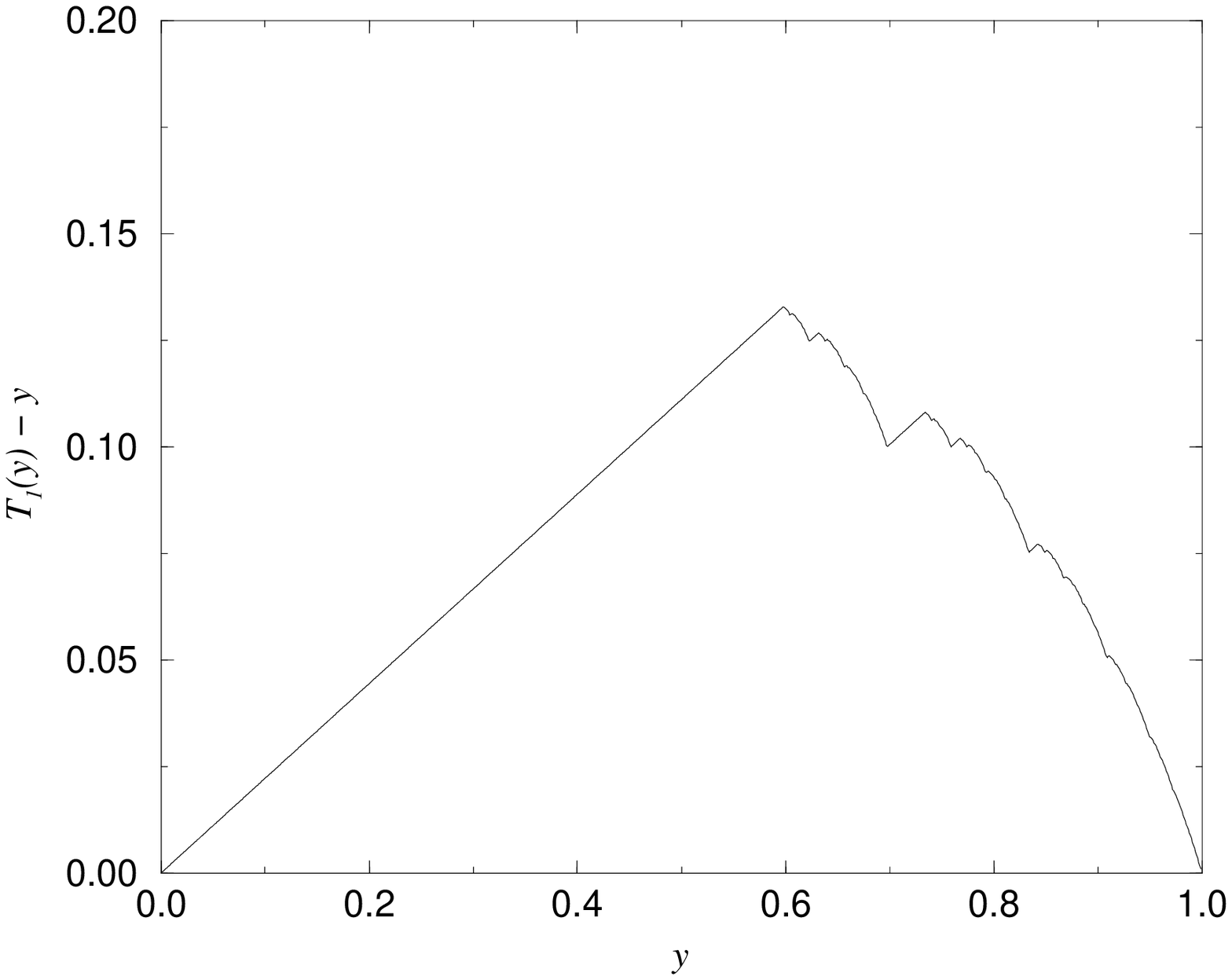}}
\caption{}
\end{figure}

\newpage

\begin{figure}[htb]
\centerline{\psfig{figure=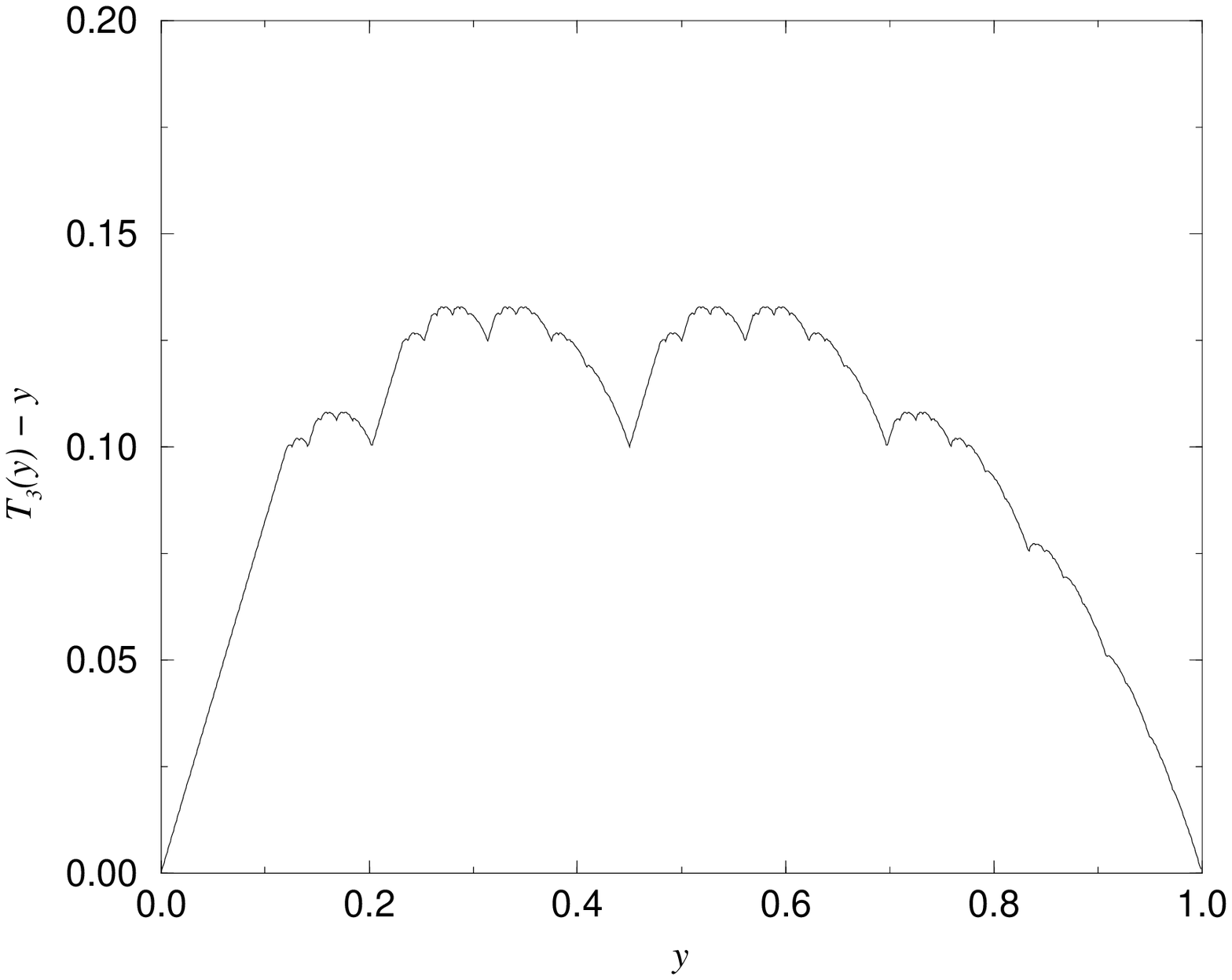}}
\caption{}
\end{figure}

\newpage

\begin{figure}[htb]
\centerline{\psfig{figure=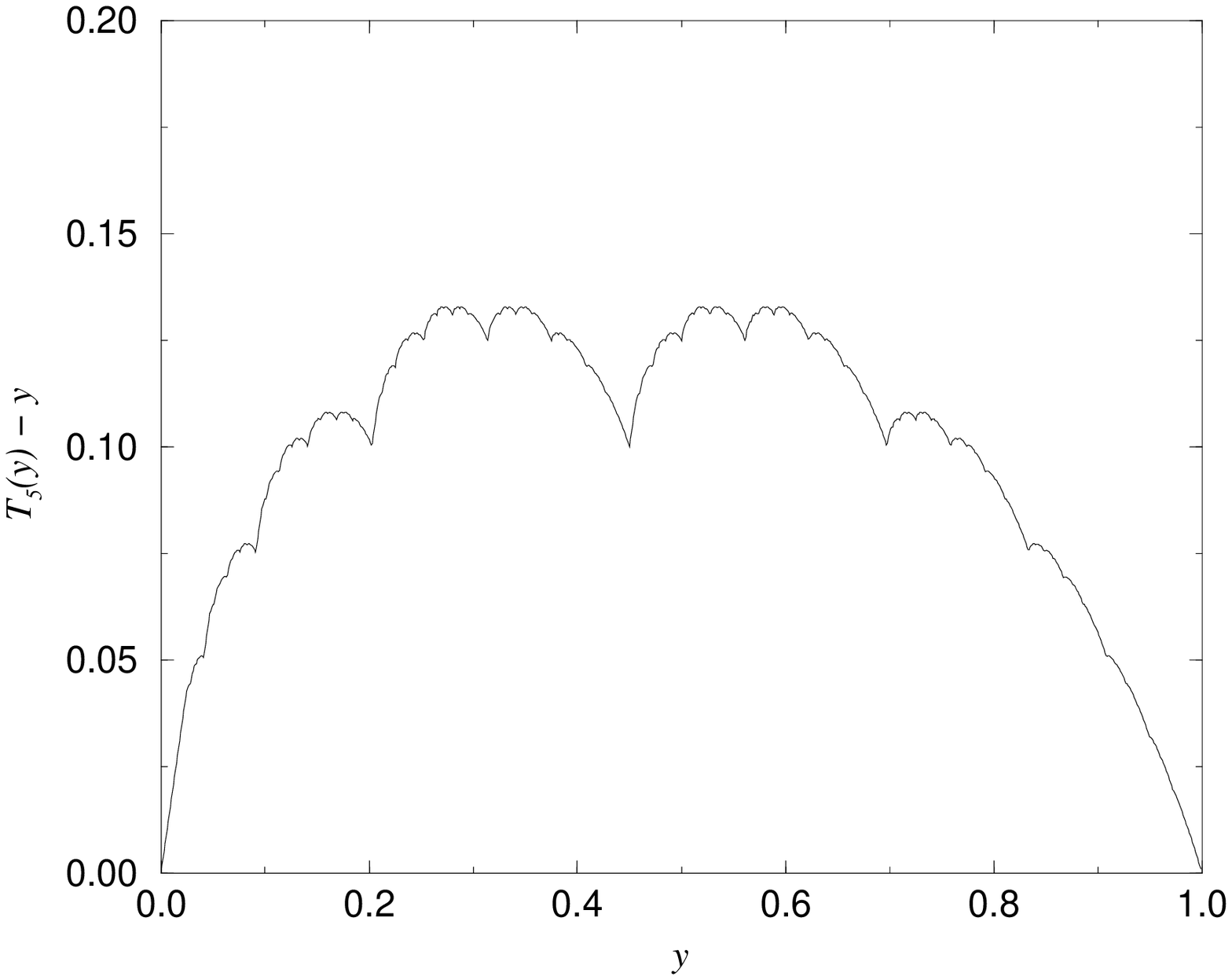}}
\caption{}
\end{figure}

\newpage

\begin{figure}[htb]
\centerline{\psfig{figure=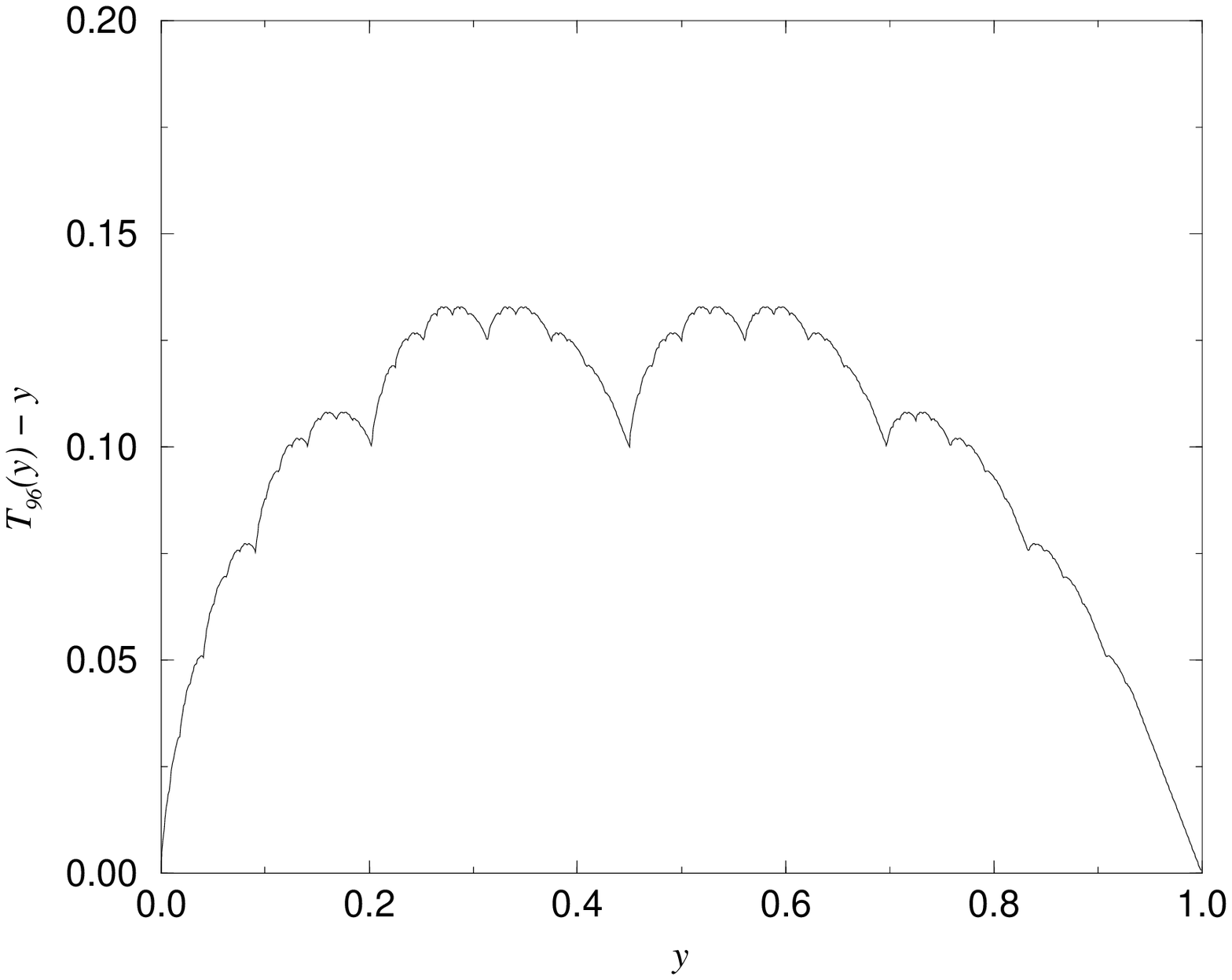}}
\caption{}
\end{figure}

\newpage

\begin{figure}[htb]
\centerline{\psfig{figure=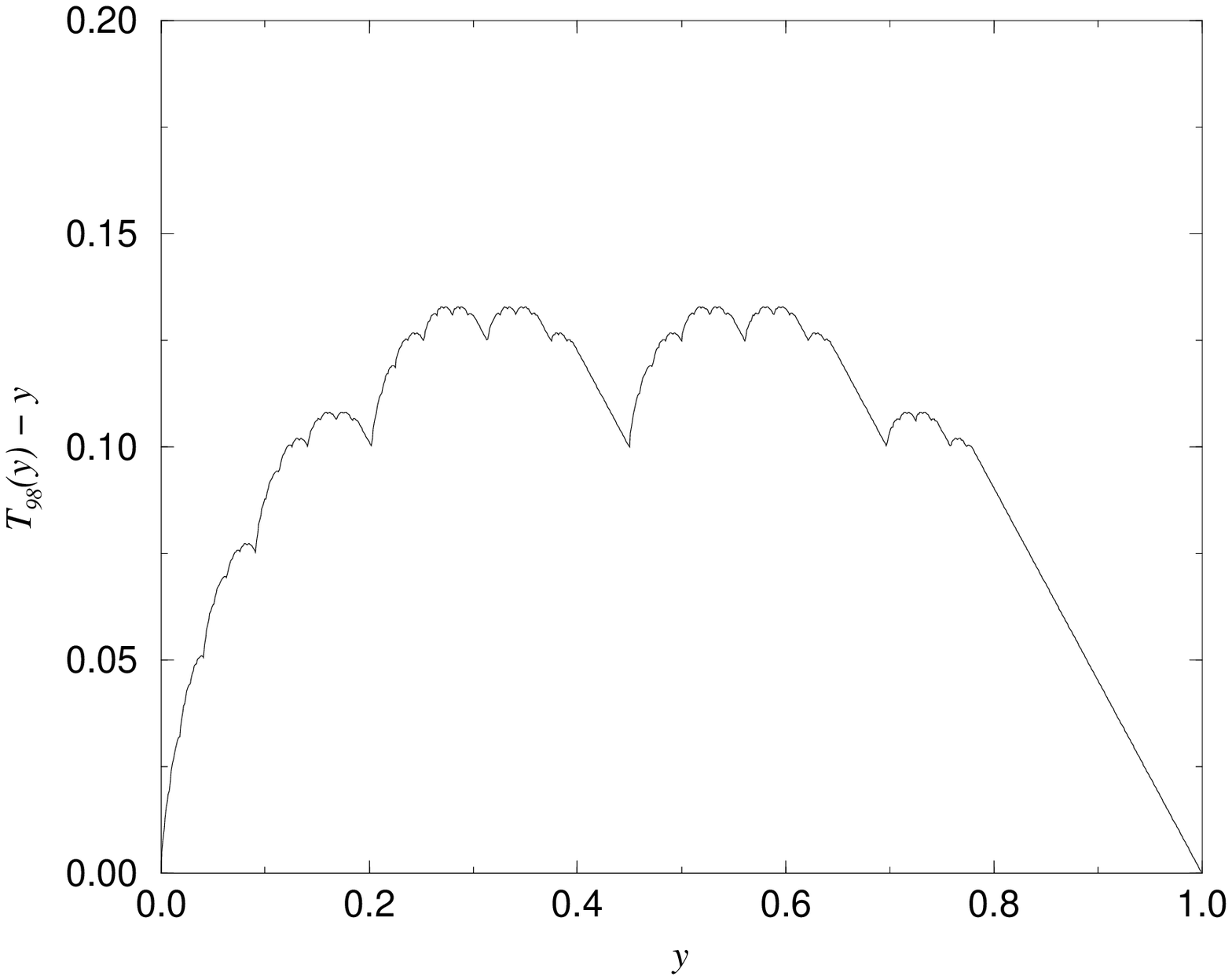}}
\caption{}
\end{figure}

\newpage

\begin{figure}[htb]
\centerline{\psfig{figure=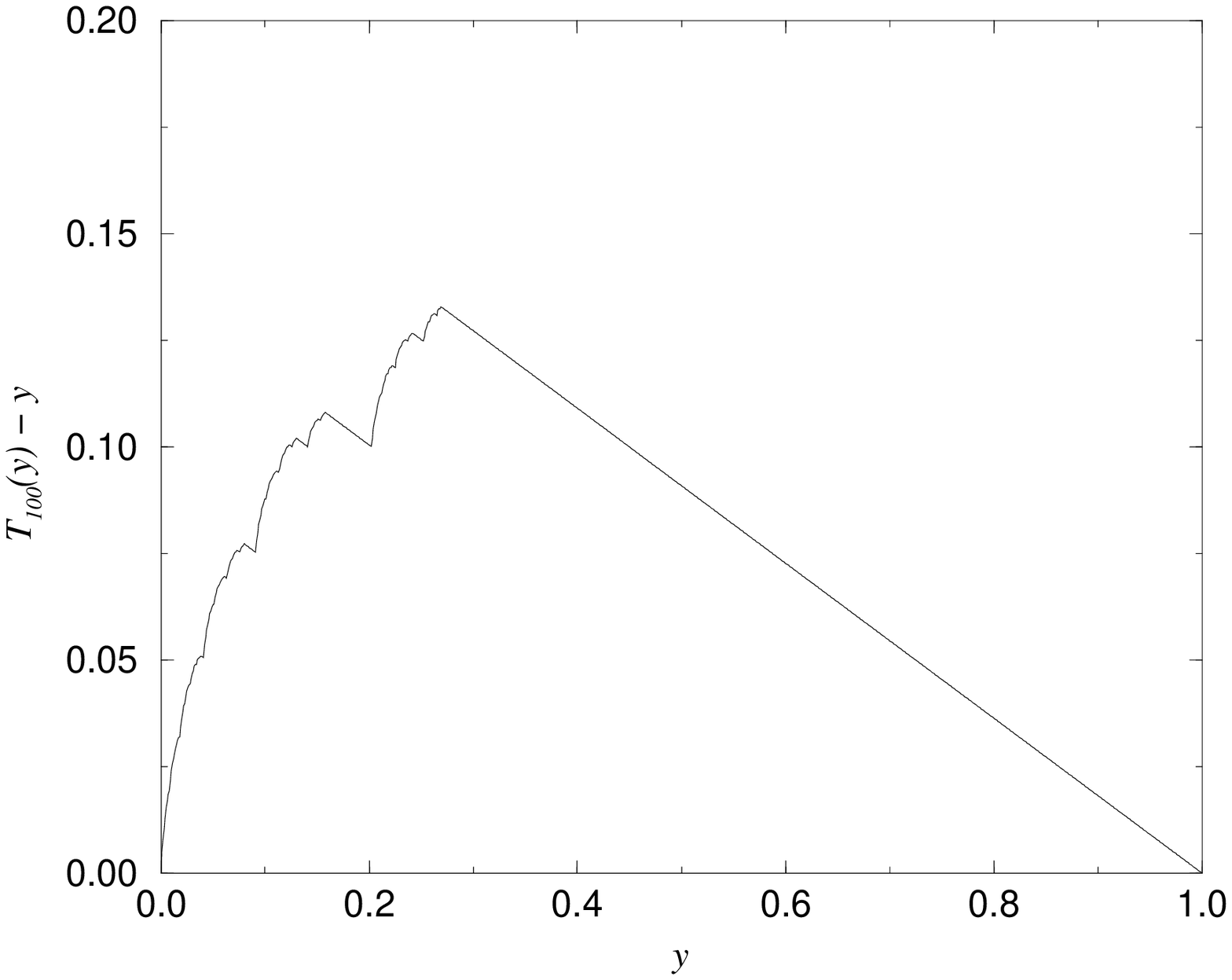}}
\caption{}
\end{figure}

\end{document}